\documentclass{caosp}
\usepackage{graphicx}

\def\I{\rm{\scriptsize I}}

\begin{document}
\pubyear{2004}
\volume{34}
\firstpage{1}
\htitle{Photometry of symbiotic stars XI.}
\hauthor{A.\,Skopal, 
         T.\,Pribulla,
         M.\,Va\v nko,
         Z. Veli\v c,
         E.\,Semkov,
         M.\,Wolf, 
    and 
         A.\,Jones}
\title{Photometry of symbiotic stars}
\subtitle{XI. EG\,And, Z\,And, BF\,Cyg, CH\,Cyg, CI\,Cyg, V1329\,Cyg, 
              TX\,CVn, AG\,Dra, RW\,Hya, AR\,Pav, AG\,Peg, 
              AX\,Per, QW\,Sge, IV\,Vir and the LMXB V934\,Her}
\author{A.\,Skopal \inst{1} 
    \and 
        T.\,Pribulla \inst{1}
    \and
        M.\,Va\v nko \inst{1} 
    \and
        Z. Veli\v c \inst{2}
    \and
        E.\,Semkov\inst{3}
    \and 
        M.\,Wolf \inst{4\,\star}\footnotetext{$^{\parbox{0.5cm}{$\star$}}
                             $Visiting Astronomer, San Pedro Observatory}
    \and 
        A.\,Jones \inst{5}
} 
\institute{
          \lomnica
       \and
          BUDA - Observatory, {\softL}. \v{S}t\'ura 16/22-16,
          018\,61 Belu\v{s}a, The Slovak Republic
       \and
          Institute of Astronomy, Bulgarian Academy of Sciences, 
          Tsarigradsko shose Blvd. 72, Sofia 1784, Bulgaria
       \and
          Astronomical Institute, Charles University Prague, CZ-180 00 
          Praha 8, \mbox{V Hole\v{s}ovi\v{c}k\'ach} 2, The Czech Republic 
       \and
          Carter Observatory, PO Box 2909, Wellington 1, New Zealand
         }
\date{December 15, 2003}

\maketitle

\begin{abstract}
We present new photometric observations of EG\,And, Z\,And, 
BF\-Cyg, CH\,Cyg, CI\,Cyg, V1329\,Cyg, TX\,CVn, AG\,Dra, 
RW\,Hya, AG\,Peg, AX\,Per, IV\,Vir and the peculiar M giant 
V934\,Her, which were made in the standard 
Johnson $UBV(R)$ system. QW\,Sge was measured in the Kron-Cousin 
$B,~V,~R_{\rm C},~I_{\rm C}$ system and for AR\,Pav we present its 
new visual estimates. The current issue gathers observations 
of these objects to December 2003. The main results can be 
summarized as follows: 
{\bf EG\,And}: 
The primary minimum in the $U$ light curve (LC) occurred at 
the end of 2002. A 0.2 -- 0.3\,mag brightening in $U$ was 
detected in the autumn of 2003. 
{\bf Z\,And}: 
At around August 2002 we detected for the first time a minimum, 
which is due to eclipse of the active object by the red giant.
Measurements from 2003.3 are close to those of a quiescent phase. 
{\bf BF\,Cyg}: 
In February 2003 a short-term flare developed in the LC.
A difference in the depth of recent minima was detected. 
{\bf CH\,Cyg}:
This star was in a quiescent phase at a rather bright state. 
A shallow minimum occurred at $\sim$\,JD~2\,452\,730, close 
to the position of the inferior conjunction of the giant 
in the inner binary of the triple-star model of CH\,Cyg. 
{\bf CI\,Cyg}:
Our observations cover the descending branch of a broad minimum. 
{\bf TX\,CVn}:
At/around the beginning of 2003 the star entered a bright stage
containing a minimum at $\sim$\,JD~2\,452\,660.
{\bf AG\,Dra}:
New observations revealed two eruptions, which peaked in 
October 2002 and 2003 at $\sim$\,9.3 in $U$. 
{\bf AR\,Pav}:
Our new visual estimates showed a transient disappearance
of a wave-like modulation in the star's brightness between 
the minima at epochs E = 66 and E = 68 and its reappearance. 
{\bf AG\,Peg}:
Our measurements from the end of 2001 showed rather complex 
profile of the LC. 
{\bf RW\,Hya}:
Observations follow behaviour of the wave-like variability of 
quiet symbiotics. 
{\bf AX\,Per}:
In May 2003 a 0.5\,mag flare was detected following a rapid
decrease of the light to a minimum. 
{\bf QW\,Sge}:
CCD observations in $B~,V,~R_{\rm C},~I_{\rm C}$ bands cover 
a period from 1994.5 to 2003.5. An increase in the star's 
brightness by about 1\,mag was observed in all passbands in 1997.
Less pronounced brightening was detected in 1999/2000. 
{\bf V934\,Her}:
Our observations did not show any larger variation in the optical 
as a reaction to its X-ray activity. 
\end{abstract}

\section{Introduction}

The symbiotic stars are currently understood as interacting 
binary systems consisting of a cool giant and a hot compact 
star, which is in most cases a white dwarf. Typical orbital 
periods are between 1 and 3 years, but they can be significantly 
larger. The mass loss from the giant represents the primary 
condition for appearance of the symbiotic phenomenon. A part 
of the material lost by the giant is transferred to the more 
compact companion via accretion from the stellar wind or 
Roche-lobe overflow. This process generates a very hot 
($T_{\rm h} \approx\,10^5$\,K) 
  and luminous 
($L_{\rm h} \approx\,10^2 - 10^4\,\rm L_{\odot}$) 
  source of radiation.
On the basis of the way in which the generated energy is 
being liberated, we distinguish two phases of symbiotic binary. 
{\em Quiescent phases} during which the hot component releases its 
energy approximately at a constant rate and spectral distribution. 
Generally, we observe a wave-like variation in their LCs as 
a function of the orbital phase. 
During {\em active phases} the hot component radiation changes 
significantly, which leads to a 2-3\,mag brightening of the 
object in the optical. A common feature of active phases is 
a high-velocity mass ejection. 

Generally, the hot radiation ionizes a fraction of the neutral 
circumbinary material, which gives rise to a strong nebular emission.  
This component of radiation is physically displaced from the hot 
star and its optically thick part can be very complex in its shape. 
In addition, its location and shaping in the binary depend on 
the level of the activity. 
As a result we often observe unexpected variation 
in the LCs as, for example, flares, drops in brightness,  
the effect of eclipses and outbursts. A very interesting feature 
of variability in this respect is the effect of eclipses, which 
is very sensitive to physical displacement and radiative 
contributions of individual components in the system.
In the case that a significant fraction of radiation at 
the wavelength under consideration comes from the region which 
is subject to eclipse, a minimum in the LC is well observable. 
In the opposite case, the eclipse effect is very faint.
Therefore the eclipse effect can be observed only at specific 
brightness phases, at which the radiative contribution from 
a pseudophotosphere in the optical rivals that from the nebula.

Accordingly, to reveal the above mentioned peculiarities in 
the LCs of symbiotic stars, it requires a very careful long-term 
monitoring programme. In this paper we present the recent 
observational results of such our programme obtained during 
the period December 2001 to December 2003. We note that this 
paper continues the work of Skopal {\it et al.} (2002, hereafter 
S+02, and references therein). 

\section{Observations}

The majority of the $U,~B,~V,~R$ measurements were performed
in the standard Johnson system using single-channel photoelectric
photometers mounted in the Cassegrain foci of 0.6-m reflectors at
the Skalnat\'{e} Pleso (hereafter SP in Tables) and Star\'{a} Lesn\'{a}
observatories (SL). Values in tables represent means of the whole 
observing cycle. Usually, a 1-hour cycle contained about 10 to 20 
individual differences between the target and the comparison. 
This approach reduced the {\em inner} uncertainty of such the means 
to $\sim\,0.020$, $\sim\,0.005$, $\sim\,0.005$ and $\sim\,0.005$\,mag 
in the $U$, $B$, $V$ and $R$ filter, respectively. Larger 
uncertainties (about 0.1\,mag) during some nights are marked in 
tables by ':'. Further details about observation procedure are 
given in Skopal {\it et al.} (1990). 

Some observations in the $B$ and $V$ bands were made with 
the 50/70/172\,cm Schmidt telescope of the National Astronomical 
Observatory Rozhen, Bulgaria (R). Other details about utilities 
and treatment of observations as well as standard stars were 
already presented in S+02. 

The $U,~B,~V$ observations of RW\,Hya and IV\,Vir were carried out 
at the San Pedro M\'artir Observatory, Baja California, Mexico (M), 
in 2003 April. Also in this case further details are given in S+02. 

Observations of QW\,Sge were performed by one of us (ZV) at his 
private station Belu\v sa near Pova\v{z}sk\'a Bystrica (PB) with 
a Newton 180/700 telescope equipped with a CCD camera based on 
the Texas Instruments chip TC~211 (from 01/01/2002 the chip was 
changed to TC-237B). A set of $B,~V,~R,~I$ filters for a modified 
Johnson-Kron-Cousins system was used. All frames were dark 
subtracted and flat fielded. Transformation to the international 
system was made by measuring the standard stars in the star-cluster 
M\,67. Magnitudes of comparison stars were obtained in the same 
way as described in Hric{\it et al.} (1996). Influence of a few 
arcsec distant companion (F0\,V star reddened with $E_{\rm B-V}$ 
= 0.20, $U$ = 13.84, $B$ = 13.59, $V$ = 13.14, $R$ = 12.53 and 
$I$ = 11.52, Munari, Buson (1991)) has not been subtracted. 

In addition, 2\,800 visual magnitude estimates of AR\,Pav were 
obtained during 1982.2 -- 2003.9 by one of us (AJ) with a private 
12".5 f/5 reflector. Other details concerning observations of AR\,Pav 
can be found in Skopal {\it et al.} 2001). 

\section{Results}

\subsection{EG\,And}

We measured EG\,And (HD\,4174, BD+39\,167) with respect to 
HD\,4143 (SAO\,63\-173, BD+37\,2318). To obtain magnitudes 
in $B$ and $V$ we used the standard star 
HD\,3914 ($V$ = 7.00, $B-V$ = 0.44) and conversion between 
both stars, HD4143 -- HD3914 = 4.640, 2.722 and 1.563 in the $U$,
$B$ and $V$ bands, respectively (Hric {\it et al.} 1991).

The data are compiled in Table~1. Figure~1 shows recent 
observations in $U$ and $B$. At/around JD\,2\,452\,600 
(November 2002) the primary minimum occurred in the $U$-LC. 
It is relatively narrow in profile with respect to those 
previously observed (e.g. at $\sim$\,JD\,2\,451\,160, see 
Fig.~1). A broad wave-like variation is less pronounced 
and, in addition, a 0.2 -- 0.3\,mag brightening in $U$ was 
detected by the latest observations in the autumn of 2003. 
These changes mean that the nebular component of radiation 
decreased and, instead, a pseudophotosphere with a more 
significant contribution in the optical was created
around the central star. In such a case the light 
in $U$ should be rather of stellar nature. This view 
should be confirmed by spectroscopic observations. 
%
%
\begin{figure}
\centering
\begin{center}
\resizebox{\hsize}{!}{\includegraphics[angle=-90]{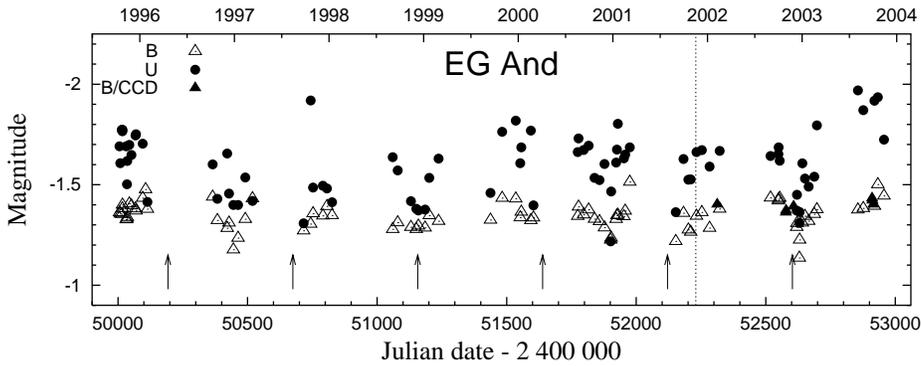}}
\caption{
 The $U$ and $B$ LCs of EG\,And. Arrows mark positions of the 
 primary minima. New data are plotted to the right of the 
 vertical dotted line.
}
\label{f1}
\end{center}
\end{figure}
%
\begin{table}[p!th]
\scriptsize
\begin{center}
\caption{$U$ and $B$ observations of EG\,And}
\begin{tabular}{lccccccc}
\hline
\hline
Date & JD~24... & Phase$^{\star}$ & $\Delta U$ & $B$ & $V$ &$\Delta R$ 
& Obs \\
\hline
 Nov 18, 01 & 52232.417 & 0.232 & -1.662 &  8.818 &  7.191 & -1.480 &SP  \\
 Dec 09, 01 & 52253.381 & 0.275 & -1.672 &  8.800 &  7.182 & -1.453 &SP  \\
 Jan 08, 02 & 52283.288 & 0.337 & -1.590 &  8.878 &  7.253 & -1.400 &SP  \\
 Feb 06, 02 & 52312.277 & 0.397 & --     &  8.760 &  7.210 & --     & R  \\
 Feb 16, 02 & 52322.257 & 0.418 & -1.668 &  8.783 &  7.148 & -1.495 &SP  \\
 Aug 31, 02 & 52518.484 & 0.825 & -1.643 &  8.728 &  7.078 & -1.584 &SP  \\
 Sep 30, 02 & 52548.464 & 0.887 & -1.652 &  8.742 &  7.097 & -1.567 &SP  \\
 Oct 02, 02 & 52549.522 & 0.890 & -1.686 &  8.735 &  7.097 & -1.548 &SP  \\
 Oct 06, 02 & 52553.615 & 0.898 & -1.619 &  8.729 &  7.084 & -1.589 &SP  \\
 Oct 29, 02 & 52577.465 & 0.948 & --     &  8.800 &  7.220 & --     & R  \\
 Oct 30, 02 & 52578.413 & 0.950 & --     &  8.790 &  7.240 & --     & R  \\
 Nov 28, 02 & 52607.345 & 0.010 & --     &  8.770 &  7.240 & --     & R  \\
 Dec 10, 02 & 52619.351 & 0.035 & -1.371 &  8.874 &  7.220 & -1.453 &SP  \\
 Dec 11, 02 & 52620.346 & 0.037 & -1.450 &  8.854 &  7.213 & -1.460 &SP  \\
 Dec 20, 02 & 52629.368 & 0.055 & -1.310 &  9.027 &  7.393 & -1.325 &SP  \\
 Dec 21, 02 & 52630.408 & 0.057 & -1.364 &  8.937 &  7.303 & -1.442 &SP  \\
 Jan 01, 03 & 52641.346 & 0.080 & -1.606 &  8.852 &  7.219 & -1.449 &SP  \\
 Jan 11, 03 & 52651.316 & 0.101 & -1.531 &  8.819 &  7.189 & -1.467 &SP  \\
 Jan 25, 03 & 52665.380 & 0.130 & -1.490 &  8.844 &  7.198 & -1.478 &SP  \\
 Feb 16, 03 & 52687.254 & 0.175 & -1.540 &  8.809 &  7.174 & -1.489 &SP  \\
 Feb 26, 03 & 52697.249 & 0.196 & -1.795 &  8.786 &  7.164 & -1.507 &SP  \\
 Aug 04, 03 & 52855.559 & 0.525 & -1.969 &  8.787 &  7.171 & -1.505 &SP  \\
 Aug 24, 03 & 52876.487 & 0.568 & -1.871 &  8.780 &  7.144 & -1.526 &SP  \\
 Sep 27, 03 & 52910.448 & 0.638 & --     &  8.730 &  7.200 & --     & R  \\
 Sep 28, 03 & 52911.396 & 0.640 & --     &  8.740 &  7.170 & --     & R  \\
 Oct 02, 03 & 52915.447 & 0.649 & --     &  8.760 &  7.200 & --     & R  \\
 Oct 06, 03 & 52919.471 & 0.657 & -1.918 &  8.769 &  7.140 & -1.521 &SP  \\
 Oct 19, 03 & 52932.475 & 0.684 & -1.935 &  8.661 &  7.007 & -1.665 &SP  \\
 Nov 12, 03 & 52956.229 & 0.733 & -1.724 &  8.718 &  7.049 &  --    &SL  \\
\hline
\hline
\end{tabular}
\end{center}
  $JD_{\rm Min} = 2\,446\,336.7 + 482\times E$ (Skopal~1997)
\normalsize
\end{table}

\subsection{Z\,And}

This star (HD\,221650) was measured with respect to the comparison 
SAO\,53150 (BD+47\,4192; $V$ = 8.99, $B - V$ = 0.41, $U - B$ = 0.14, 
$V - R$ = 0.16). Other details are the same as in S+02. Results 
are given in Table~2. 

Figure~2 shows our photometric observations from 2000 covering 
a recent major outburst with the beginning at September 2000 
and a maximum in December of that year (Skopal {\it et al.} 2000\,a). 
Then a gradual decrease lasted up to June 2002. At around August 2002 
we detected for the first time a minimum, which was due 
to eclipse of the active object by the red giant (Skopal 2003). 
This result suggests a very high inclination of the orbital plane 
of Z\,And. Further observations indicated an increase in 
the star's brightness to the spring of 2003. The latest 
measurements from 2003.3 are close to those of the quiescent phase. 
This indicates a low level of activity and/or just a maximum of 
the wave-like variability at the orbital phase 0.5 (see Fig.~2, 
Table~2). Further observations will demonstrate that. 
%
%
%
\begin{figure}
\centering
\begin{center}
\resizebox{\hsize}{!}{\includegraphics[angle=-90]{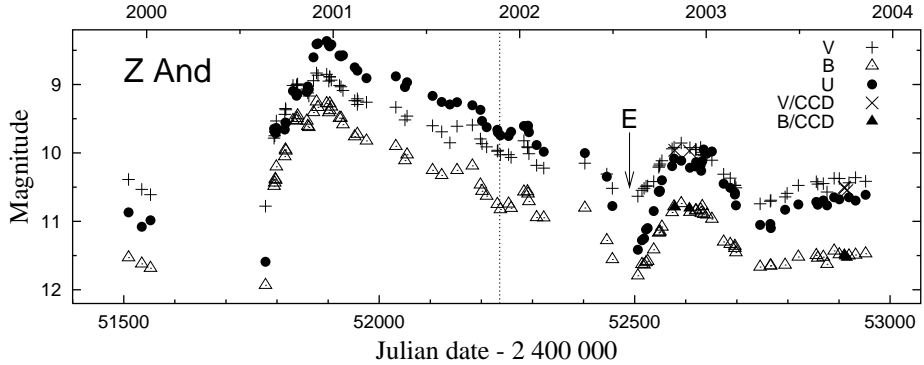}}
\caption{
 The $U,~B,~V$ photometry of Z\,And covering the recent active
 phase. The eclipse of the active component by the red giant is
 denoted by {\sf E}.
}
\label{f2}
\end{center}
\end{figure}
%
%
\begin{table}[p!bh]
\scriptsize
\begin{center}
\caption{$U,~B,~V,~R$ observations of Z\,And}
\begin{tabular}{lccccccc}
\hline
\hline
Date & JD~24... & Phase$^{\star}$  & $U$ &  $B$   &  $V$   &  $R$   & Obs \\
\hline
 Nov 23, 01 & 52237.255 & 0.653 &  9.743 & 10.827 & 10.027 & --     & SL \\
 Dec 09, 01 & 52253.338 & 0.674 &  9.753 & 10.745 & 10.025 &  9.027 &SP  \\
 Dec 14, 01 & 52258.200 & 0.681 &  9.691 & 10.808 & 10.065 & --     & SL \\
 Jan 08, 02 & 52283.247 & 0.714 &  9.605 & 10.568 &  9.822 &  8.933 &SP  \\
 Jan 15, 02 & 52290.357 & 0.723 &  9.642 & 10.564 &  9.910 &  8.959 &SP  \\
 Jan 16, 02 & 52291.235 & 0.724 &  9.601 & 10.597 &  9.922 &  8.977 &SP  \\
 Jan 18, 02 & 52293.244 & 0.727 &  9.696 & 10.710 & 10.011 &  9.040 &SP  \\
 Feb 02, 02 & 52308.224 & 0.747 &  9.88: & 10.94: & 10.18: & --     & SL \\
 Feb 16, 02 & 52322.221 & 0.765 &  9.981 & 10.948 & 10.222 &  9.167 &SP  \\
 May 08, 02 & 52402.538 & 0.871 & 10.002 & 10.805 & 10.150 &  9.179 &SP  \\
 Jun 19, 02 & 52445.457 & 0.928 & 10.346 & 11.279 & 10.309 & --     & SL \\
 Jul 01, 02 & 52456.508 & 0.942 & 10.775 & 11.557 & 10.517 &  9.306 &SP  \\
 Aug 19, 02 & 52506.451 & 0.008 & 11.414 & 11.794 & 10.633 &  9.420 &SP  \\
 Aug 27, 02 & 52514.396 & 0.019 & 11.275 & 11.629 & 10.545 & --     & SL \\
 Sep 01, 02 & 52518.539 & 0.024 & 11.250 & 11.637 & 10.516 &  9.323 &SP  \\
 Sep 04, 02 & 52522.386 & 0.029 & 11.118 & 11.601 & 10.485 & --     & SL \\
 Sep 08, 02 & 52525.507 & 0.033 & 11.098 & 11.582 & 10.482 & --     & SL \\
 Sep 19, 02 & 52537.383 & 0.049 & 10.848 & 11.415 & 10.418 &  9.267 &SP  \\
 Sep 29, 02 & 52547.301 & 0.062 & 10.561 & 11.170 & 10.207 & --     & SL \\
 Sep 30, 02 & 52548.426 & 0.064 & 10.568 & 11.163 & 10.188 &  9.081 &SP  \\
 Oct 01, 02 & 52549.483 & 0.065 & 10.552 & 11.151 & 10.161 &  9.070 &SP  \\
 Oct 06, 02 & 52553.520 & 0.070 & 10.399 & 11.083 & 10.113 &  9.032 &SP  \\
 Oct 25, 02 & 52573.225 & 0.096 & 10.19: & 10.87: &  9.94: &  8.940 &SP  \\
 Oct 29, 02 & 52577.341 & 0.102 & 10.082 & 10.784 &  9.918 & --     &SL  \\
 Oct 30, 02 & 52578.402 & 0.103 & --     & 10.790 &  9.960 & --     & R  \\
 Nov 12, 02 & 52591.216 & 0.120 & 10.114 & 10.738 &  9.854 & --     &SL  \\
 Nov 28, 02 & 52607.333 & 0.141 & --     & 10.810 &  9.970 & --     & R  \\
 Nov 29, 02 & 52608.181 & 0.143 & 10.215 & 10.866 &  9.922 &  8.813 &SP  \\
 Dec 10, 02 & 52619.317 & 0.157 & 10.197 & 10.844 &  9.947 &  8.843 &SP  \\
 Dec 11, 02 & 52620.313 & 0.159 & 10.133 & 10.842 &  9.951 &  8.853 &SP  \\
 Dec 20, 02 & 52629.325 & 0.170 & 10.176 & 10.829 &  9.918 &  8.857 &SP  \\
 Dec 21, 02 & 52630.370 & 0.172 & 10.260 & 10.880 & 10.020 &  8.900 &SP  \\
 Dec 23, 02 & 52632.325 & 0.174 & 10.120 & 10.781 &  9.989 &  8.857 &SP  \\
 Dec 26, 02 & 52635.260 & 0.178 &  9.949 & 10.897 & 10.040 & --     &SL  \\
 Jan 01, 03 & 52641.292 & 0.186 & 10.009 & 10.905 & 10.024 &  8.858 &SP  \\
 Jan 11, 03 & 52651.261 & 0.199 &  9.978 & 10.966 & 10.106 &  8.929 &SP  \\
 Feb 03, 03 & 52674.216 & 0.230 & 10.450 & 11.300 & 10.310 &  9.090 &SP  \\
 Feb 16, 03 & 52687.223 & 0.247 & 10.509 & 11.335 & 10.367 &  9.091 &SP  \\
 Feb 25, 03 & 52696.283 & 0.259 & 10.580 & 11.363 & 10.472 & --     &SL  \\
 Feb 25, 03 & 52696.240 & 0.259 & 10.608 & 11.393 & 10.474 &  9.169 &SP  \\
 Feb 27, 03 & 52698.251 & 0.261 & 10.765 & 11.456 & 10.509 &  9.195 &SP  \\
 Apr 16, 03 & 52745.576 & 0.324 & 11.051 & 11.668 & 10.743 & --     &SL  \\
 May 06, 03 & 52765.543 & 0.350 & 11.037 & 11.652 & 10.705 &  9.387 &SP  \\
 May 07, 03 & 52766.540 & 0.352 & 11.097 & 11.648 & 10.691 &  9.382 &SP  \\
 Jun 03, 03 & 52794.474 & 0.388 & 10.829 & 11.638 & 10.643 & --     &SL  \\
 Jun 07, 03 & 52798.399 & 0.394 &  --    &  --    & 10.6:: &  9.30: &SP  \\
 Jun 29, 03 & 52820.416 & 0.423 & 10.752 & 11.520 & 10.475 &  9.193 &SP  \\
 Aug 04, 03 & 52855.525 & 0.469 & 10.716 & 11.497 & 10.420 &  9.145 &SP  \\
 Aug 06, 03 & 52858.487 & 0.473 & 10.756 & 11.535 & 10.454 &  9.160 &SP  \\
 Aug 17, 03 & 52869.400 & 0.487 & 10.695 & 11.522 & 10.447 &  9.154 &SP  \\
 Aug 24, 03 & 52876.452 & 0.497 & 10.766 & 11.627 & 10.571 & --     &SP  \\
 Sep 08, 03 & 52890.588 & 0.515 & 10.648 & 11.431 & 10.369 &  9.078 &SP  \\
 Sep 18, 03 & 52900.560 & 0.529 & 10.677 & 11.486 & 10.374 &  9.093 &SP  \\
 Sep 27, 03 & 52910.434 & 0.542 & --     & 11.490 & 10.500 & --     & R  \\
 Sep 28, 03 & 52911.384 & 0.543 & --     & 11.510 & 10.510 & --     & R  \\
 Oct 02, 03 & 52915.408 & 0.548 & --     & 11.530 & 10.550 & --     & R  \\
 Oct 06, 03 & 52919.425 & 0.553 & 10.647 & 11.503 & 10.436 &  9.150 &SP  \\
 Oct 19, 03 & 52932.441 & 0.571 & 10.697 & 11.492 & 10.361 &  9.078 &SP  \\
 Nov 08, 03 & 52952.222 & 0.597 & 10.612 & 11.472 & 10.414 & --     &SL  \\
\hline
\hline
\end{tabular}
\end{center}
  $JD_{\rm Min} = 2\,414\,625.2 + 757.5\times E$ (Skopal~1998)
\normalsize
\end{table}

\subsection{BF\,Cyg}

The photometric measurements of BF\,Cyg are given in Table~3. 
Stars 
 HD\,183650 
 ($V$ = 6.96, $B-V$ = 0.71, $U-B$ = 0.34, $V-R$ = 0.56 )
and 
 BD+30\,3594 
 ($V$ = 9.54, $B-V$ = 1.20, $U-B$ = 1.70)
were used as the comparison ~and ~check, respectively. 

Figure~3 shows the $U,B,V$ LCs from 1999.6. A few V/CCD 
observations available from VSNET (made by O. Pejcha) were 
added for comparison. The periodic wave-like 
variation in the optical continuum reflects a quiescent phase of 
this star. However, the profile of the LC is not a simple sinusoid 
through the orbital cycle. It differs in many details from cycle 
to cycle. Generally, such behaviour reflects a complex shape and 
variation of the nebula in the binary (Skopal 2001). For example, 
the minimum around JD~2\,452\,160 was rather flat for about 110 
days (best seen in $B$). The current minimum ($\sim$\,JD~2\,452\,930) 
seems to follow the same behaviour, but very close to its mid 
($\varphi$\,=\,0.986, see Table~3) we detected a dip in $U$ 
(= 13.08) and $B$ (= 13.52), which represents the lowest brightness 
ever observed for BF\,Cyg. We note that shortly after this 
detection, on 15th October 2003 ($\varphi$\,=\,0.996), BF\,Cyg fell
down in its brightness under the limit of detection within our 
devices (data quality was poor due to high background). 
The latest observations follow those taken just prior 
to this dip (Fig.~3). Additional peculiarity in the LC developed 
in February 2003 in a form of a short-term flare. This transient 
brightening lasted for about 1 month (our observations did not 
record its accurate profile) and was most pronounced in $B$ 
($\Delta B \sim$\,0.7\,mag). It occurred at the orbital phase 0.68, 
very close in the phase to that observed for AX\,Per (see below, 
Sect.~3.12). The nature of such brightening is not well 
understood. It differs from flares/outbursts currently observed 
for other symbiotics (e.g. AG\,Dra, here in Fig.~7), amplitude 
of which is largest in $U$. 
%
%
%
\begin{figure}
\centering
\begin{center}
\resizebox{\hsize}{!}{\includegraphics[angle=-90]{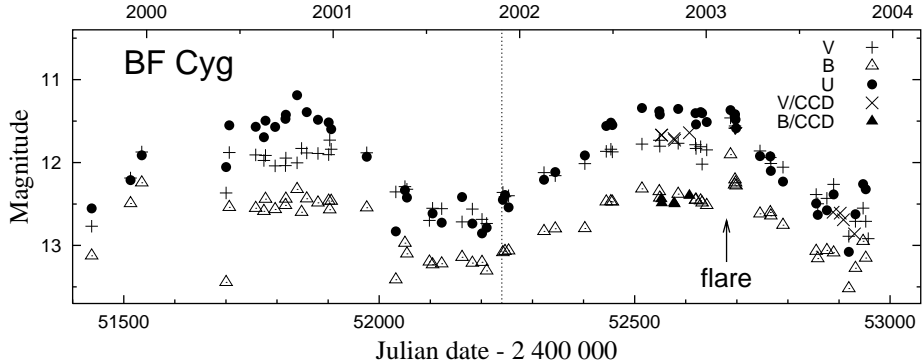}}
\caption{
 The $UBV$ LCs of BF\,Cyg. A small flare developed in February 2003. 
}
\label{f3}
\end{center}
\end{figure}
%
%
\begin{table}[!ht]
\scriptsize
\begin{center}
\caption{$U,~B,~V,~R$ observations of BF\,Cyg. A few points were added 
from the VSNET database.}
\begin{tabular}{lccccccc}
\hline
\hline
Date & JD~24... & Phase$^{\star}$ & $U$ & $B$ & $V$ & $\Delta R$ & Obs \\
\hline
 Nov 28, 01 & 52242.257 & 0.092 & 12.450 & 13.085 & 12.360 &  4.751 &SP  \\
 Dec 02, 01 & 52246.214 & 0.097 & 12.392 & 13.071 & 12.411 & --     & SL \\
 Dec 09, 01 & 52253.195 & 0.106 & 12.541 & 13.065 & 12.400 &  4.684 &SP  \\
 Feb 17, 02 & 52322.630 & 0.198 & 12.207 & 12.831 & 12.121 &  4.382 &SP  \\
 Mar 11, 02 & 52344.616 & 0.227 & 12.116 & 12.798 & 12.159 &  4.449 &SP  \\
 May 07, 02 & 52402.485 & 0.303 & 11.915 & 12.795 & 12.013 &  4.175 &SP  \\
 Jun 18, 02 & 52444.482 & 0.359 & 11.560 & 12.474 & 11.865 &  4.057 &SP  \\
 Jun 27, 02 & 52453.456 & 0.371 & 11.520 & 12.470 & 11.840 &  4.140 &SP  \\
 Jun 30, 02 & 52456.462 & 0.375 & 11.548 & 12.476 & 11.872 &  4.106 &SP  \\
 Aug 27, 02 & 52514.357 & 0.451 & 11.343 & 12.318 & 11.778 & --     & SL \\
 Sep 30, 02 & 52548.250 & 0.496 & 11.381 & 12.349 & 11.732 &  3.888 &SP  \\
 Oct 01, 02 & 52549.321 & 0.497 & 11.423 & 12.428 & 11.803 &  3.956 &SP  \\
 Oct 04, 02 & 52552.309 & 0.501 & --     & 12.440 & 11.670 & --     & R  \\
 Oct 05, 02 & 52553.234 & 0.502 & --     & 12.490 & 11.670 & --     & R  \\
 Oct 29, 02 & 52577.244 & 0.534 & --     & 12.490 & 11.710 & --     & R  \\
 Oct 30, 02 & 52578.206 & 0.535 & --     & 12.500 & 11.730 & --     & R  \\
 Nov 06, 02 & 52585.244 & 0.545 & 11.354 & 12.381 & 11.768 &  3.945 &SP  \\
 Nov 28, 02 & 52607.179 & 0.574 & --     & 12.400 & 11.640 & --     & R  \\
 Dec 10, 02 & 52619.180 & 0.590 & 11.405 & 12.454 & 11.786 &  3.989 &SP  \\
 Dec 11, 02 & 52620.192 & 0.591 & 11.540 & 12.451 & 11.830 &  4.045 &SP  \\
 Dec 20, 02 & 52629.188 & 0.603 & 11.393 & 12.461 & 11.810 &  4.110 &SP  \\
 Dec 23, 02 & 52632.189 & 0.607 & 11.405 & 12.487 & 12.020 &  1.910$^{\dagger}$ &SP  \\
 Jan 01, 03 & 52641.199 & 0.619 & 11.511 & 12.516 & 11.848 &  4.091 &SP  \\
 Feb 17, 03 & 52687.681 & 0.680 & 11.368 & 11.905 & 11.463 &  1.584$^{\dagger}$ &SP  \\
 Feb 25, 03 & 52695.650 & 0.691 & 11.435 & 12.270 & 11.585 & --     & SL \\
 Feb 26, 03 & 52696.597 & 0.692 & 11.418 & 12.207 & 11.559 &  3.848 &SP  \\
 Feb 27, 03 & 52697.599 & 0.693 & 11.481 & 12.245 & 11.590 &  3.848 &SP  \\
 Feb 28, 03 & 52698.622 & 0.694 & 11.587 & 12.281 & 11.579 &  3.899 &SP  \\
 Apr 15, 03 & 52745.473 & 0.756 & 11.923 & 12.618 & 11.860 & --     & SL \\
 May 05, 03 & 52765.459 & 0.783 & 11.926 & 12.603 & 11.940 &  4.166 &SP  \\
 May 06, 03 & 52766.476 & 0.784 & 12.101 & 12.642 & 12.011 &  4.233 &SP  \\
 May 30, 03 & 52790.494 & 0.816 & 12.230 & 12.755 & 12.057 &  4.289 &SP  \\
 Aug 03, 03 & 52855.448 & 0.902 & 12.495 & 13.070 & 12.386 &  4.656 &SP  \\
 Aug 06, 03 & 52858.338 & 0.905 & 12.631 & 13.158 & 12.498 &  4.718 &SP  \\
 Aug 24, 03 & 52876.346 & 0.929 & 12.575 & 13.060 & 12.433 &  4.658 &SP  \\
 Sep 05, 03 & 52888.437 & --    & --     & --     & 12.60  &  -- & VSNET \\
 Sep 06, 03 & 52889.369 & 0.946 & 12.38: & 13.09: & 12.26: &  4.49: &SP  \\
 Sep 19, 03 & 52902.495 & --    & --     & --    & 12.61  &  -- & VSNET \\
 Sep 25, 03 & 52908.521 & --    & --     & --    & 12.69  &  -- & VSNET \\
 Oct 06, 03 & 52919.306 & 0.986 & 13.08: & 13.52: & 12.890 &  4.99: &SP  \\
 Oct 15, 03 & 52928.292 & 0.036 &$\downarrow$&$\downarrow$& 12.918 & -- & SL \\
 Oct 17, 03 & 52930.401 & --    & --     & --     & 12.86 &  -- & VSNET \\
 Oct 19, 03 & 52932.296 & 0.003 & 12.624 & 13.275 & 12.710 &  4.981 &SP  \\
 Nov 03, 03 & 52947.256 & 0.023 & 12.26: & 12.95: & 12.55: &  4.737 &SP  \\
 Nov 08, 03 & 52952.247 & 0.029 & 12.322 & 13.152 & 12.709 &  --    &SL  \\
\hline
\hline
\end{tabular}
\end{center}
 $^{\star}$ $Min = JD\,2\,415\,065 + 757.3\times E$
           (Pucinskas 1970) \\
 $^{\dagger}$ $\Delta R$ = BF\,Cyg - BD+30\,3594 \\
 $\downarrow$ target within the backround
\normalsize
\end{table}

\subsection{CH\,Cyg}

Our new photometry of CH\,Cyg is listed in Table~4. 
Stars 
 HD\,183123 (SAO\,48428, $V$ = 8.355, $B-V$ = 0.478, $U-B$ = -- 0.031,
 $V-R$ = 0.312)
and 
 HD\,182691
 (SAO\,31623, $V$ = 6.525, $B-V$ = -- 0.078, $U-B$ = -- 0.24, 
$V-R$ = 0)
were used as a comparison and a check star, respectively. 
In addition, we also measured the nearby star SAO~31628 
(BD+49$^{\circ}$2997) to examine its variability suggested 
by Sokoloski, Stone (2000). However, our type of measurements 
was not suitable for such a short-period variable 
($P_{\rm orb}$ = 3.74783 d). We were able to confirm only 
the position of the primary minimum according to the ephemeris 
suggested by Sokoloski, Stone (2000). Further observations 
are required to determine the whole LC-profile. 

Figure~4 shows the recent photometric observations covering 
the last 1998-00 activity including the eclipse at the outer, 
14.5-year period, orbit. Our new observations indicated 
evolution in LCs, which is similar to that occurred after 
the previous active phases, in 1970, 1987 and 1996.5 
(see Fig.~1 of Eyres {\it et al.} 2002). This is characterized 
by a 750-day wave-like variation in LCs and rather bright 
magnitudes ($V \sim\,7 - 7.5$, $B \sim\,9 - 9.5$ and $U \approx\,10$). 
The colour indices are typical for a quiescent phase of this star. 
At $\sim\,$JD~2\,452\,730 a shallow minimum occurred in the $UBV$ LCs. 
Its position is very close to that predicted according to the 
ephemeris, $Min = JD\,2\,445\,888 + 756\times E$ (Skopal 1995). 
%
%
%
\begin{figure}
\centering
\begin{center}
\resizebox{\hsize}{!}{\includegraphics[angle=-90]{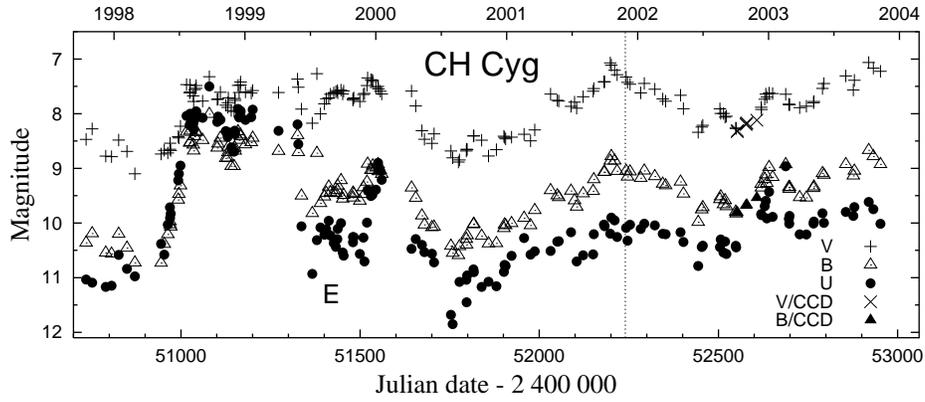}}
\caption{
 The $UBV$ LCs of CH\,Cyg. 
}
\label{f4}
\end{center}
\end{figure}
%
%
\begin{table}[!ht]
\scriptsize
\begin{center}
\caption{$U,~B,~V,~R$ observations of CH\,Cyg}
\begin{tabular}{lccccccc}
\hline
\hline
Date & JD~24... & Phase$^{\star}$ & $U$ & $B$ & $V$ & $\Delta R$ & Obs \\
\hline
 Nov 28, 01 & 52242.308 & 0.405 & 10.091 &  9.053 &  7.328 & -1.006 &SP  \\
 Dec 02, 01 & 52246.179 & 0.410 & 10.324 &  9.158 &  7.428 & --     & SL \\
 Dec 09, 01 & 52253.291 & 0.420 & 10.032 &  9.064 &  7.461 & -0.972 &SP  \\
 Jan 08, 02 & 52283.196 & 0.459 & 10.109 &  9.183 &  7.620 & -0.846 &SP  \\
 Jan 18, 02 & 52293.202 & 0.472 & 10.014 &  9.048 &  7.447 & -0.977 &SP  \\
 Feb 17, 02 & 52322.517 & 0.511 & 10.044 &  9.150 &  7.552 & -0.848 &SP  \\
 Mar 11, 02 & 52344.568 & 0.540 & 10.199 &  9.287 &  7.732 & -0.733 &SP  \\
 Mar 19, 02 & 52352.563 & 0.551 & 10.206 &  9.318 &  7.772 & -0.692 &SP  \\
 Apr 29, 02 & 52394.400 & 0.606 & 10.169 &  9.249 &  7.663 & -0.734 &SP  \\
 May 07, 02 & 52402.435 & 0.617 & 10.349 &  9.468 &  7.909 & -0.551 &SP  \\
 Jun 17, 02 & 52443.455 & 0.671 & 10.787 &  9.987 &  8.341 & --     & SL \\
 Jun 27, 02 & 52453.405 & 0.684 & 10.440 &  9.757 &  8.244 & -0.326 &SP  \\
 Jun 30, 02 & 52456.416 & 0.688 & 10.416 &  9.717 &  8.214 & -0.346 &SP  \\
 Aug 17, 02 & 52504.325 & 0.752 & 10.442 &  9.569 &  7.913 & --     & SL \\
 Aug 20, 02 & 52506.500 & 0.755 & 10.303 &  9.520 &  7.996 & -2.034 &SP  \\
 Aug 27, 02 & 52514.307 & 0.765 & 10.542 &  9.663 &  7.996 & --     & SL \\
 Aug 31, 02 & 52518.430 & 0.770 & 10.342 &  9.590 &  8.063 & -1.983 &SP  \\
 Sep 04, 02 & 52522.344 & 0.776 & 10.568 &  9.700 &  8.047 & --     & SL \\
 Sep 30, 02 & 52548.343 & 0.810 & 10.419 &  9.798 &  8.263 & -1.765 &SP  \\
 Oct 01, 02 & 52549.448 & 0.811 & 10.447 &  9.832 &  8.272 & -1.785 &SP  \\
 Oct 04, 02 & 52552.351 & 0.815 & --     &  9.820 &  8.330 & --     & R  \\
 Oct 05, 02 & 52553.217 & 0.816 & --     &  9.810 &  8.310 & --     & R  \\
 Oct 29, 02 & 52577.209 & 0.848 & --     &  9.680 &  8.200 & --     & R  \\
 Oct 30, 02 & 52578.224 & 0.850 & --     &  9.680 &  8.180 & --     & R  \\
 Oct 31, 02 & 52579.190 & 0.851 & --     &  9.670 &  8.170 & --     & R  \\
 Nov 28, 02 & 52607.162 & 0.888 & --     &  9.540 &  8.120 & --     & R  \\
 Dec 10, 02 & 52619.223 & 0.904 &  9.837 &  9.290 &  7.884 & -2.094 &SP  \\
 Dec 11, 02 & 52620.233 & 0.905 &  9.857 &  9.359 &  7.923 & -2.038 &SP  \\
 Dec 20, 02 & 52629.285 & 0.917 &  9.692 &  9.198 &  7.735 & -2.194 &SP  \\
 Dec 23, 02 & 52632.226 & 0.921 &  9.596 &  9.117 &  7.643 & -2.241 &SP  \\
 Dec 26, 02 & 52635.191 & 0.925 &  9.916 &  9.277 &  7.698 & --     & SL \\
 Jan 01, 03 & 52641.238 & 0.933 &  9.429 &  8.983 &  7.622 & -2.325 &SP  \\
 Jan 11, 03 & 52651.209 & 0.946 &  9.887 &  9.166 &  7.633 & -2.341 &SP  \\
 Feb 17, 03 & 52687.620 & 0.994 &  8.962 &  8.932 &  7.645 & -2.307 &SP  \\
 Feb 26, 03 & 52696.665 & 0.006 &  9.924 &  9.350 &  7.827 & -2.217 &SP  \\
 Feb 27, 03 & 52697.640 & 0.007 & 10.014 &  9.381 &  7.835 & -2.205 &SP  \\
 Feb 28, 03 & 52698.584 & 0.009 &  9.872 &  9.347 &  7.835 & -2.212 &SP  \\
 Mar 28, 03 & 52726.609 & 0.046 & 10.209 &  9.533 &  7.890 & --     & SL \\
 Apr 15, 03 & 52745.401 & 0.071 & 10.211 &  9.542 &  7.864 & --     & SL \\
 May 05, 03 & 52765.414 & 0.097 & 10.010 &  9.368 &  7.802 & -2.225 &SP  \\
 May 06, 03 & 52766.424 & 0.098 &  9.974 &  9.338 &  7.778 & -2.235 &SP  \\
 May 30, 03 & 52790.450 & 0.130 &  9.824 &  9.095 &  7.538 & -2.426 &SP  \\
 Jun 03, 03 & 52794.423 & 0.135 &  9.998 &  9.125 &  7.450 & --     & SL \\
 Aug 03, 03 & 52855.403 & 0.216 &  9.798 &  8.928 &  7.310 & -2.617 &SP  \\
 Aug 24, 03 & 52876.414 & 0.244 &  9.870 &  9.138 &  7.568 & -2.420 &SP  \\
 Aug 27, 03 & 52879.331 & 0.248 &  9.710 &  9.050 &  7.390 & -2.420 &SP  \\
 Oct 06, 03 & 52919.350 & 0.301 &  9.615 &  8.671 &  7.065 & -2.737 &SP  \\
 Oct 19, 03 & 52932.406 & 0.318 &  9.748 &  8.785 &  7.167 & -2.757 &SP  \\
 Nov 08, 03 & 52952.277 & 0.344 & 10.015 &  8.928 &  7.225 & --     & SL \\
\hline
\hline
\end{tabular}
\end{center}
 $^{\star}$ $Min = JD\,2\,445\,888 + 756\times E$
           (Skopal 1995) \\
\normalsize
\end{table}
%

\subsection{CI\,Cyg}

The photometric measurements of CI\,Cyg are given in Table~5.
Stars
 HD\,226107 
 (SAO~68948; $V$ = 8.55, $B-V$ = -- 0.04, $U-B$ = -- 0.33) 
and
 HD\,226041
 (SAO~68923; $V$ = 8.60, $B-V$ = 0.35, spectrum F\,5)
were used as the comparison and check, respectively.

We started monitoring of CI\,Cyg from September 30, 2002 (the 
orbital phase $\varphi$ = 0.53, see Table~5, Fig.~5). At that 
time it was around its maximum, which lasted to about March 2003 
($\varphi \sim$\,0.75) with a small diminution in $U$. Then 
the star's brightness was decreasing to a minimum at 
$\varphi \sim$\,0 by about of 1\,mag in $U$ and $B$ and by about 
0.5\,mag in $V$. Such behaviour is typical for a quiescent 
phase of symbiotic stars, and for CI\,Cyg was recently nicely 
demonstrated by Dmitrienko (2000) by a large series of 
the $UBVRI$ observations during 1996 -- 1999. In addition to 
the orbitally related variability, the $V$-LC displays 
40 -- 60-day variations with the amplitude of 0.2 -- 0.4\,mag, 
which is also seen well in the averaged visual LC (Fig.~5). 
This type of variability was originally revealed by 
Belyakina, Prokofieva (1991), who ascribed it to the cool 
giant in CI\,Cyg. 
%
%
%
\begin{figure}
\centering
\begin{center}
\resizebox{\hsize}{!}{\includegraphics[angle=-90]{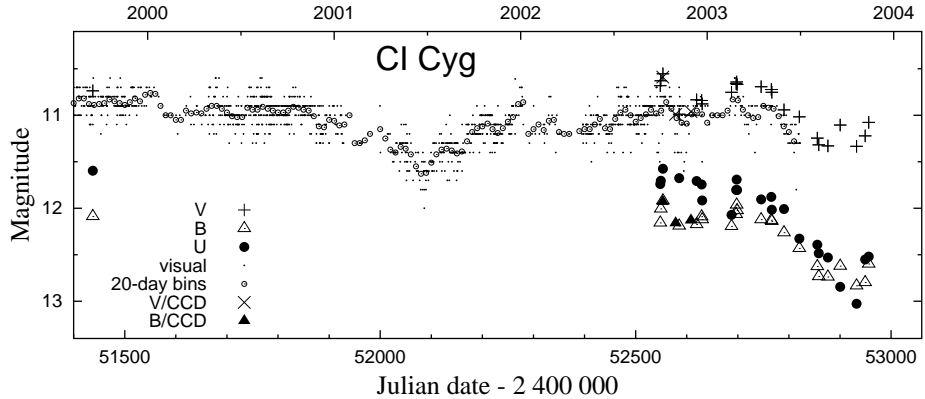}}
\caption{
 The $UBV$ LCs of CI\,Cyg.
}
\label{f5}
\end{center}
\end{figure}
%
%
\begin{table}
\scriptsize
\begin{center}
\caption{$U,~B,~V$ observations of CI\,Cyg}
\begin{tabular}{lccccccc}
\hline
\hline
Date & JD~24... & Phase$^{\star}$ & $U$ & $B$ & $V$ & $\Delta R$ & Obs \\
\hline
 Sep 15, 99 & 51437.474 & 0.227 & 11.597 & 12.088 & 10.738 &  0.779 &SP  \\
 Sep 30, 02 & 52548.296 & 0.526 & 11.740 & 12.158 & 10.682 &  0.674 &SP  \\
 Oct 01, 02 & 52549.387 & 0.527 & 11.704 & 12.007 & 10.630 &  0.664 &SP  \\
 Oct 05, 02 & 52553.254 & 0.531 &  --    & 11.930 & 10.590 & --     & R  \\
 Oct 05, 02 & 52553.422 & 0.532 & 11.576 & 11.919 & 10.555 &  0.582 &SP  \\
 Oct 30, 02 & 52578.238 & 0.561 &  --    & 12.160 & 10.990 & --     & R  \\
 Nov 06, 02 & 52585.357 & 0.569 & 11.676 & 12.190 & 10.989 &  0.885 &SP  \\
 Nov 29, 02 & 52608.184 & 0.596 &  --    & 12.130 & 10.960 & --     & R  \\
 Dec 10, 02 & 52619.277 & 0.609 & 11.708 & 12.174 & 10.834 &  0.774 &SP  \\
 Dec 20, 02 & 52629.231 & 0.620 & 11.745 & 12.091 & 10.841 &  0.819 &SP  \\
 Dec 21, 02 & 52630.180 & 0.621 & 11.92: & 12.12: & 10.88: &  0.92: &SP  \\
 Feb 17, 03 & 52687.653 & 0.689 & 12.071 & 12.195 & 10.753 &  0.744 &SP  \\
 Feb 26, 03 & 52696.633 & 0.699 & 11.803 & 12.065 & 10.659 &  0.701 &SP  \\
 Feb 27, 03 & 52697.666 & 0.700 & 11.692 & 11.965 & 10.642 &  0.695 &SP  \\
 Feb 28, 03 & 52698.661 & 0.701 & 11.804 & 12.024 & 10.669 &  0.720 &SP  \\
 Apr 16, 03 & 52745.530 & 0.756 & 11.905 & 12.122 & 10.694 & --     &SL  \\
 May 06, 03 & 52765.507 & 0.780 & 11.878 & 12.129 & 10.723 &  0.780 &SP  \\
 May 07, 03 & 52766.509 & 0.781 & 12.016 & 12.141 & 10.753 &  0.807 &SP  \\
 May 31, 03 & 52790.529 & 0.809 & 12.008 & 12.263 & 10.942 &  0.974 &SP  \\
 Jun 29, 03 & 52820.379 & 0.844 & 12.327 & 12.432 & 11.018 &  1.028 &SP  \\
 Aug 03, 03 & 52855.492 & 0.885 & 12.392 & 12.626 & 11.246 &  1.247 &SP  \\
 Aug 06, 03 & 52858.405 & 0.888 & 12.484 & 12.734 & 11.318 &  1.266 &SP  \\
 Aug 24, 03 & 52876.381 & 0.909 & 12.530 & 12.739 & 11.329 &  1.310 &SP  \\
 Sep 17, 03 & 52900.466 & 0.937 & 12.845 & 12.622 & 11.105 &  1.112 &SP  \\
 Oct 19, 03 & 52932.347 & 0.975 & 13.026 & 12.833 & 11.334 &  1.256 &SP  \\
 Nov 03, 03 & 52949.195 & 0.994 & 12.551 & 12.798 & 11.222 &  1.182 &SP \\
 Nov 12, 03 & 52956.191 & 0.003 & 12.520 & 12.599 & 11.077 & --     & SL\\
\hline
\hline
\end{tabular}
\end{center}
  $\star$ JD$_{\rm Min}$ = 2\,411\,902 + 855.25$\times$E
\normalsize
\end{table}
%

\subsection{V1329\,Cyg}

Our observations of the symbiotic nova V1329\,Cyg (HBV\,475) are
given in Table~6. In this paper we present only the CCD observations 
made at the Rozhen Observatory. 
The stars 
BD+35\,4290 ($V$ = 10.34, $B-V$ = 1.07, $U-B$ = 0.88) and
BD+35\,4294 ($V$ = 10.16, $B-V$ = 1.07)
were used as the comparison and check, respectively. 
%
\begin{table}
\scriptsize
\begin{center}
\caption{CCD $B$ and $V$ observations of V1329\,Cyg from the Rozhen
Observatory. Three points were added from the VSNET database for comparison.}
\begin{tabular}{lccccc}
\hline
\hline
Date   &  JD~24... & Phase$^{\star}$&  $B$&  $V$ & Obs \\
\hline
 Oct 04, 02 & 52552.362 & 0.955  & 14.86 & 13.94 & R  \\
 Oct 05, 02 & 52553.270 & 0.956  & 14.86 & 13.96 & R  \\
 Oct 29, 02 & 52577.267 & 0.981  & 14.78 & 13.92 & R  \\
 Oct 30, 02 & 52578.250 & 0.983  & 14.76 & 13.91 & R  \\
 Nov 28, 02 & 52607.196 & 0.013  & 14.62 & 13.80 & R  \\
 Apr 03, 03 & 52732.609 & 0.144  & 14.15 & 13.28 & R  \\
 May 03, 03 & 52762.529 & 0.175  & 14.08 & 13.20 & R  \\
 May 05, 03 & 52765.499 & 0.178  & 14.09 & 13.22 & R  \\
 Aug 06, 03 & 52858.433 & 0.275  &  --   & 13.17 & VSNET\\
 Aug 15, 03 & 52867.458 & 0.284  &  --   & 12.98 & VSNET\\
 Aug 28, 03 & 52880.408 & 0.298  &  --   & 12.79 & VSNET\\ 
 Sep 27, 03 & 52910.246 & 0.329  & 13.66 & 12.95 & R  \\
 Sep 28, 03 & 52911.345 & 0.330  & 13.66 & 12.93 & R  \\
 Oct 02, 03 & 52915.371 & 0.334  & 13.69 & 13.00 & R  \\
\hline
\hline
\end{tabular}
\end{center}
$^{\star}$ $JD_{\rm eclipse} = 2\,427\,687 + 958.0 \times E$
           (Schild, Schmid 1997)
\normalsize
\end{table}
%

\subsection{TX\,CVn}

The results of our photometric measurements of TX\,CVn 
(HD\,63173, BD+37\-2318) are depicted in Table~7.
Stars 
BD+382374 
(SAO\,63223, $V$ = 9.36, $B-V$ = 0.30, $U-B$ = 0.03) 
and
HD\,111113 
(SAO\,63189, $V$ = 9.18, $B-V$ = 0.38, $U-B$ = -- 0.07),
were used as a comparison and a check star, respectively. 

Figure~6 shows our recent $U,~B,~V$ measurements. With respect to 
the evolution in the historical LC, TX\,CVn still remains at 
a high level of its activity ($B \sim$\,10.5), while at low stages 
the photographic LC was at $m_{\rm pg} \sim$\,11.6 (see Fig.~1 of 
Skopal {\it et al.} 2000\,b). In addition, we indicated two 
brightenings on our recent $U,~B,~V$ LCs. First occurred at 
the end of 1996 and the second one at the beginning of 2003. 
During both we detected a minimum, which 
can be ascribed to the eclipse of the active component by its 
cool giant companion in the TX\,CVn binary. The reasons are as 
follows: (i) The minima occurred very close to the inferior 
conjunction of the giant according to solution for the spectroscopic 
orbit as proposed by Kenyon, Garcia (1989) and, (ii) both minima 
were more pronounced in $U$ than in $B$. The mid points of these 
minima (JD~2450\,477.6\,$\pm\,1.0$ and JD~2452\,660\,$\pm\,10$) 
suggest the orbital period 
\begin{center}
  $P_{\rm orb} = 198.4 \pm 0.9$ days,
\end{center}
which agrees within uncertainties with that suggested by 
Kenyon, Garcia (1989) for a circular orbit solution. 
The eclipsing nature of the observed minima suggest a high 
inclination of the orbital plane of TX\,CVn. 
%
%
%
\begin{figure}
\centering
\begin{center}
\resizebox{\hsize}{!}{\includegraphics[angle=-90]{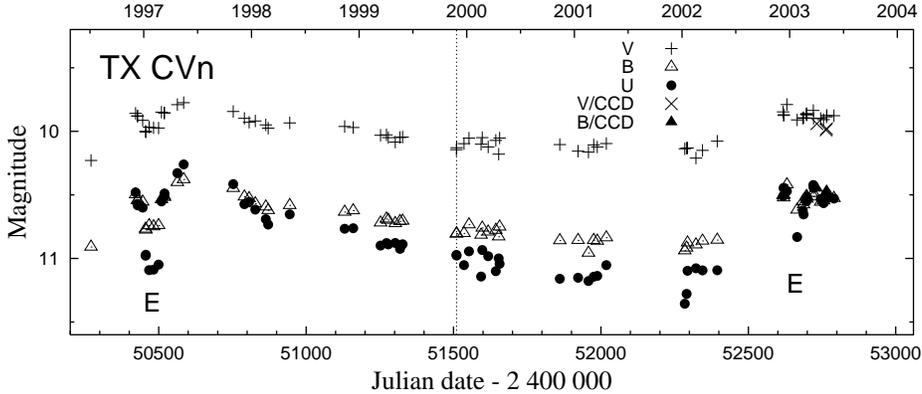}}
\caption{
 The $UBV$ LCs of TX\,CVn. Eclipses of the active star
by its giant companion are denoted by {\sf E}. 
}
\label{f6}
\end{center}
\end{figure}
%
%
\begin{table}
\scriptsize
\begin{center}
\caption{$U,~B,~V,~R$ observations of TX\,CVn}
\begin{tabular}{lccccccc}
\hline
\hline
Date & JD~24... & Phase$^{\star}$ & $U$ & $B$ & $V$ & $\Delta R$ & Obs \\
\hline
 Nov 28, 99 & 51510.664 & 0.223 & 10.977 & 10.805 & 10.129 &
 0.592$^{\dagger}$ &SP  \\
 Dec 23, 99 & 51535.688 & 0.350 & 11.053 & 10.803 & 10.097 &  0.324 &SP  \\
 Jan 08, 00 & 51552.437 & 0.434 & 10.944 & 10.735 & 10.054 &
 0.495$^{\dagger}$ &SP  \\
 Feb 18, 00 & 51593.410 & 0.641 & 11.143 & 10.816 & 10.102 & --     & SL \\
 Feb 22, 00 & 51597.420 & 0.661 & 10.934 & 10.760 & 10.050 &
 0.512$^{\dagger}$ &SP  \\
 Mar 13, 00 & 51617.372 & 0.762 & 10.982 & 10.791 & 10.123 &  0.343 &SP  \\
 Apr 08, 00 & 51643.496 & 0.894 & 11.100 & 10.778 & 10.072 & --     & SL \\
 Apr 18, 00 & 51653.320 & 0.944 & 11.000 & 10.830 & 10.180 &
 0.610$^{\dagger}$ &SP  \\
 Apr 21, 00 & 51656.350 & 0.959 & 11.042 & 10.752 & 10.053 & --     & SL \\
 Nov 12, 00 & 51860.668 & 0.991 & 11.160 & 10.860 & 10.105 &  0.344 &SP  \\
 Jan 13, 01 & 51922.550 & 0.304 & 11.152 & 10.858 & 10.156 &  0.368 &SP  \\
 Feb 16, 01 & 51957.462 & 0.480 & 11.178 & 10.958 & 10.163 &  0.381 &SP  \\
 Mar 06, 01 & 51975.403 & 0.570 & 11.144 & 10.857 & 10.107 &  0.308 &SP  \\
 Mar 18, 01 & 51987.404 & 0.631 & 11.137 & 10.864 & 10.123 &  0.338 &SP  \\
 Apr 18, 01 & 52018.340 & 0.787 & 11.05: & 10.837 & 10.096 &  0.311 &SP  \\
 Jan 10, 02 & 52284.544 & 0.132 & 11.356 & 10.939 & 10.138 &  0.323 &SP  \\
 Jan 16, 02 & 52290.543 & 0.162 & 11.279 & 10.917 & 10.133 &  0.314 &SP  \\
 Jan 19, 02 & 52293.637 & 0.178 & 11.09: & 10.877 & 10.132 &  0.336 &SP  \\
 Feb 16, 02 & 52322.362 & 0.323 & 11.077 & 10.893 & 10.211 &  0.351 &SP  \\
 Mar 10, 02 & 52344.483 & 0.435 & 11.094 & 10.863 & 10.150 &  0.342 &SP  \\
 Apr 29, 02 & 52394.323 & 0.686 & 11.093 & 10.855 & 10.078 &  0.258 &SP  \\
 Dec 09, 02 & 52617.608 & 0.814 & 10.510 & 10.504 &  9.847 &  0.088 &SP  \\
 Dec 11, 02 & 52619.557 & 0.824 & 10.447 & 10.523 &  9.875 &  0.115 &SP  \\
 Dec 12, 02 & 52620.506 & 0.829 & 10.445 & 10.501 &  9.870 &  0.061 &SP  \\
 Dec 22, 02 & 52630.560 & 0.879 & 10.47: & 10.42: &  9.79: &  0.04: &SP  \\
 Jan 26, 03 & 52665.529 & 0.056 & 10.832 & 10.619 &  9.910 &  0.132 &SP  \\
 Feb 13, 03 & 52684.466 & 0.152 & 10.623 & 10.554 &  9.897 &  0.131 &SP  \\
 Feb 16, 03 & 52687.368 & 0.166 & 10.654 & 10.578 &  9.894 &  0.134 &SP  \\
 Feb 25, 03 & 52696.451 & 0.212 & 10.551 & 10.518 &  9.866 &  0.104 &SP  \\
 Feb 26, 03 & 52697.404 & 0.217 & 10.548 & 10.512 &  9.861 &  0.111 &SP  \\
 Feb 27, 03 & 52698.426 & 0.222 & 10.515 & 10.532 &  9.871 &  0.116 &SP  \\
 Mar 21, 03 & 52720.365 & 0.333 & 10.423 & 10.487 &  9.836 &  0.095 &SP  \\
 Mar 22, 03 & 52721.334 & 0.338 & 10.444 & 10.516 &  9.898 &  0.111 &SP  \\
 Apr 02, 03 & 52732.419 & 0.394 &  --    & 10.450 &  9.940 & --     & R  \\
 Apr 15, 03 & 52745.321 & 0.459 & 10.533 & 10.557 &  9.899 & --     & SL \\
 Apr 25, 03 & 52755.320 & 0.509 & 10.565 & 10.547 &  9.908 & --     & SL \\
 May 03, 03 & 52763.367 & 0.550 &  --    & 10.490 &  9.99  & --     & R  \\
 May 05, 03 & 52765.326 & 0.560 &  --    & 10.470 &  9.98  & --     & R  \\
 May 05, 03 & 52765.333 & 0.560 & 10.490 & 10.520 &  9.887 &  0.130 &SP  \\
 May 06, 03 & 52766.345 & 0.565 & 10.529 & 10.524 &  9.875 &  0.118 &SP  \\
 May 30, 03 & 52790.374 & 0.686 & 10.529 & 10.527 &  9.877 &  0.106 &SP  \\
\hline
\hline
\end{tabular}
\end{center}
 $^{\star}$ $JD_{\rm sp. conj.} = 2\,445\,130.45 + 198\times E$
           (Kenyon, Garcia 1989) \\
  $^{\dagger}$ $\Delta R$ = TX\,CVn - HD\,111113
\normalsize
\end{table}
%

\subsection{AG\,Dra}

Our measurements of AG\,Dra are summarized in Table~8. 
Stars 
BD+67\,925 
(SAO\,16952, $V$ = 9.88, $B-V$ = 0.56, $U-B$ = -- 0.04) 
and 
BD+67\,923 
(SAO\,16935, $V$ = 9.46, $B-V$ = 1.50, $U-B$ = 1.89) 
were used as the comparison and check, respectively. 

Figure~7 shows the $U,B,V$ LCs covering 
the recent declining part of the massive outburst, which began 
in 1994 July. The LC from 1994 was characterized by numerous 
eruptions (see Fig.~6 of S+02). Between 1994.5 and 1998.5 they 
appeared regularly with a period of about 1 year. After 1998.5, 
eruptions were not so regular, they had a lower amplitude and 
were not so massive as prior to this time. New observations 
revealed two short-term eruptions, which peaked in October 2002 
and 2003, respectively, at $\sim$\,9.3 in $U$. This indicates 
that the recent activity of AG\,Dra gradually dies away. 
However, the wave-like variation, typical for a quiescent 
phase, has not been developed yet. 
%
%
%
\begin{figure}
\centering
\begin{center}
\resizebox{\hsize}{!}{\includegraphics[angle=-90]{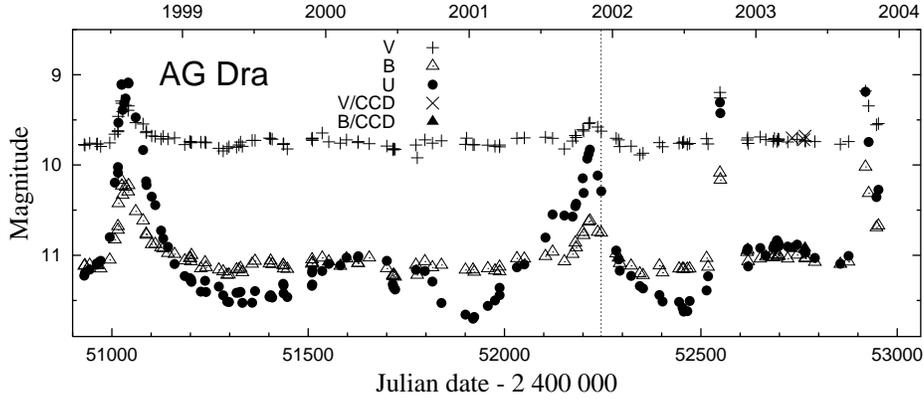}}
\caption{
 The $UBV$ LCs of AG\,Dra. 
}
\label{f7}
\end{center}
\end{figure}
%
\begin{table}
\scriptsize
\begin{center}
\caption{$U,~B,~V,~R$ observations of AG\,Dra}
\begin{tabular}{lccccccc}
\hline
\hline
Date & JD~24... & Phase$^{\star}$ & $U$ & $B$ & $V$ & $\Delta R$ & Obs \\
\hline
 Dec 02, 01 & 52246.277 & 0.675 & 10.291 & 10.753 &  9.625 & --     & SL \\
 Jan 09, 02 & 52284.493 & 0.745 & 10.948 & 10.982 &  9.704 & -0.704 &SP  \\
 Jan 16, 02 & 52290.500 & 0.756 & 11.046 & 11.029 &  9.721 & -0.690 &SP  \\
 Jan 19, 02 & 52293.594 & 0.761 & 11.172 & 11.090 &  9.796 & -0.600 &SP  \\
 Feb 16, 02 & 52322.402 & 0.814 & 11.229 & 11.117 &  9.792 & -0.645 &SP  \\
 Mar 11, 02 & 52344.535 & 0.854 & 11.341 & 11.203 &  9.888 & -0.560 &SP  \\
 Mar 19, 02 & 52352.525 & 0.868 & 11.368 & 11.223 &  9.869 & -0.561 &SP  \\
 Apr 29, 02 & 52394.365 & 0.945 & 11.440 & 11.116 &  9.753 & -0.588 &SP  \\
 May 07, 02 & 52402.397 & 0.959 & 11.512 & 11.193 &  9.798 & -0.594 &SP  \\
 Jun 18, 02 & 52444.429 & 0.036 & 11.516 & 11.150 &  9.759 & -0.631 &SP  \\
 Jun 27, 02 & 52453.367 & 0.052 & 11.581 & 11.153 &  9.752 & -0.611 &SP  \\
 Jun 30, 02 & 52456.369 & 0.057 & 11.624 & 11.154 &  9.744 & -0.635 &SP  \\
 Jul 09, 02 & 52465.452 & 0.074 & 11.619 & 11.153 &  9.771 & -0.632 &SP  \\
 Jul 15, 02 & 52471.426 & 0.085 & 11.507 & 11.149 &  9.761 & -0.623 &SP  \\
 Aug 27, 02 & 52514.499 & 0.163 & 11.390 & 11.033 &  9.709 & --     & SL \\
 Aug 31, 02 & 52518.365 & 0.170 & 11.233 & 11.131 &  9.767 & -0.589 &SP  \\
 Oct 01, 02 & 52548.500 & 0.225 &  9.307 & 10.087 &  9.196 & -0.984 &SP  \\
 Oct 02, 02 & 52549.576 & 0.227 &  9.426 & 10.167 &  9.258 & -0.932 &SP  \\
 Dec 09, 02 & 52617.643 & 0.351 & 10.947 & 10.966 &  9.723 & -0.683 &SP  \\
 Dec 10, 02 & 52619.413 & 0.354 & 10.927 & 10.943 &  9.700 & -0.722 &SP  \\
 Dec 11, 02 & 52620.418 & 0.356 & 11.12: & 11.067 &  9.761 & -0.637 &SP  \\
 Jan 11, 03 & 52651.457 & 0.412 & 10.9:: & 11.04: &  9.69: & -0.550 &SP  \\
 Jan 25, 03 & 52665.481 & 0.438 & 11.004 & 11.019 &  9.719 & -0.682 &SP  \\
 Feb 13, 03 & 52684.393 & 0.472 & 10.918 & 11.026 &  9.736 & -0.690 &SP  \\
 Feb 16, 03 & 52687.328 & 0.477 & 10.885 & 10.946 &  9.705 & -0.703 &SP  \\
 Feb 23, 03 & 52693.617 & 0.489 & 10.835 & 11.019 &  9.739 & --     & SL \\
 Feb 25, 03 & 52696.397 & 0.494 & 10.885 & 10.994 &  9.713 & -0.700 &SP  \\
 Feb 26, 03 & 52697.360 & 0.496 & 10.865 & 11.010 &  9.728 & -0.699 &SP  \\
 Feb 27, 03 & 52698.375 & 0.498 & 10.900 & 11.009 &  9.715 & -0.709 &SP  \\
 Mar 22, 03 & 52721.398 & 0.539 & 10.903 & 11.040 &  9.745 & -0.674 &SP  \\
 Apr 02, 03 & 52732.451 & 0.560 & --     & 10.93  &  9.70  & --     & R  \\
 May 02, 03 & 52762.417 & 0.614 & --     & 10.94  &  9.72  & --     & R  \\
 Apr 15, 03 & 52745.353 & 0.583 & 10.882 & 10.998 &  9.716 & --     & SL \\
 May 05, 03 & 52765.343 & 0.619 & --     & 10.92  &  9.68  & --     & R  \\
 May 05, 03 & 52765.381 & 0.619 & 10.957 & 11.029 &  9.735 & -0.691 &SP  \\
 May 06, 03 & 52766.393 & 0.621 & 10.974 & 11.039 &  9.733 & -0.681 &SP  \\
 May 30, 03 & 52790.413 & 0.665 & 11.030 & 11.076 &  9.741 & -0.681 &SP  \\
 Aug 03, 03 & 52855.364 & 0.783 & 11.094 & 11.098 &  9.770 & -0.644 &SP  \\
 Aug 24, 03 & 52876.303 & 0.821 & 11.008 & 11.074 &  9.740 & -0.664 &SP  \\
 Oct 06, 03 & 52919.387 & 0.900 &  9.189 & 10.022 &  9.181 & -1.072 &SP  \\
 Oct 14, 03 & 52927.251 & 0.914 &  9.745 & 10.314 &  9.345 & --     & SL \\
 Nov 03, 03 & 52947.198 & 0.950 & 10.355 & 10.693 &  9.555 & -0.846 &SP \\
 Nov 08, 03 & 52952.190 & 0.959 & 10.275 & 10.673 &  9.542 & --     &SL \\
\hline
\hline
\end{tabular}
\end{center}
$^{\star}$ $Min = JD\,2\,443\,629.17 + 549.73\times E$
           (G\'alis {\it et al.} 1999)
\normalsize
\end{table}
%

\subsection{RW\,Hya}

The $U,B,V$ measurements of RW\,Hya (HD\,117970) are listed 
in Table~9. The observation was carried out at the San Pedro 
Observatory during April 2003, at its orbital phase 
$\varphi \sim\,0.75$. Note that the orbital period is very 
close to just 1 year (Table~9). 
Stars 
HD\,118102 (CD-24\,10984;  $V$ = 8.944, $B-V$ = 0.528, $U-B$ = 0.105), 
HD\,117971 (CD-25\,9879; $V$ = 9.688, $B-V$ = 0.439, $U-B$ = -- 0.034) 
and 
HD\,117803 (CD-24\,10970; $V$ = 8.925, $B-V$ = 0.417, $U-B$ = -- 0.025) 
were used as standard stars, to which RW\,Hya was compared. 

Magnitudes are brighter, mainly in the $U$ band, with respect 
to our previous observations made at $\varphi \sim\,0.89$. 
This is in a good agreement with the wave-like variation as 
a function of the orbital phase suggested by visual estimates 
(see Fig.~7 of S+02). Observations at other positions of 
the binary, mainly at $\varphi \sim\,0.5$, are very desirable. 
%
%
%
%
\begin{table}[!ht]
\scriptsize
\begin{center}
\caption{$U,B,V$ observations of RW\,Hya}
\begin{tabular}{lcccccc}
\hline
\hline
Date & JD~24... & Phase$^{\star}$ & $U$ &   $B$   &   $V$  & Obs \\
\hline
 Apr 20, 03 & 52749.772 & 0.741 & 10.295 & 10.184 &  8.743 & M  \\
 Apr 20, 03 & 52749.814 & 0.741 & 10.229 & 10.159 &  8.758 & M  \\
 Apr 20, 03 & 52749.816 & 0.741 & 10.233 & 10.183 &  8.748 & M  \\
 Apr 22, 03 & 52751.814 & 0.747 & 10.314 & 10.193 &  8.739 & M  \\
 Apr 22, 03 & 52751.817 & 0.747 & 10.284 & 10.135 &  8.737 & M  \\
 Apr 23, 03 & 52752.849 & 0.750 & 10.373 & 10.174 &  8.754 & M  \\
 Apr 23, 03 & 52752.852 & 0.750 & 10.278 & 10.180 &  8.720 & M  \\
\hline
\hline
\end{tabular}
\end{center}
$^{\star}$ $JD_{\rm sp. conj.} = 2\,449\,512 + 370.4\times E$
           (Schild {\it et al.} 1996)
\normalsize
\end{table}
%

\subsection{AR\,Pav}

Figure~8 shows the visual LC from 1982 to December 2003. Some 
qualitative discussion of a major part of these data can be found 
in Skopal {\it et al.} (2000\,b, 2001). Our new visual estimates cover 
the period from the epoch 67 (1999.8, bottom panel of Fig.~8). The 
most interesting feature of LC is a transient disappearance 
of a wave-like modulation of the star's brightness as a function 
of the orbital phase for the period of just two cycles, between 
the minima at epochs E = 66 and E = 68. Also variation in 
the depth of the minima (e.g. minima at E = 61,~68 are by about 
0.3-0.4\,mag brighter than those at E = 64) and in its profile 
(see also Fig.~2 of Skopal {\it et al.} 2001) reflect a strong 
variation in both geometry and radiation of the active component 
of AR\,Pav. The latest observations from E = 68 indicate a 
follow-up transition to a high state similar to that observed 
prior to the cycle 66. Perfect agreement between the photoelectric 
photometry and our visual estimates suggests that the described 
details in the visual LC are real (see Skopal {\it et al.} 2001).  
Finally, we determined position of the recent minimum to 
Min(68) = JD~2\,452364.5\,$\pm$\,0.7. This position 
and those of 
Min(66) = JD~2\,451\,158.9\,$\pm$\,0.7 
and 
Min(67) = JD~2\,451\,762.8\,$\pm$\,0.7 
(Skopal {\it et al.} 2001), which occurred during the low stage, 
correspond to the period of only 602.8\,$\pm$\,0.3 days. This is 
by 1.65 day shorter than that given by all available mid-points. 
On the other hand, it is close to that predicted by the parabolic 
ephemeris derived by Skopal {\it et al.} (2000\,c). 
%
%
%
\begin{figure}
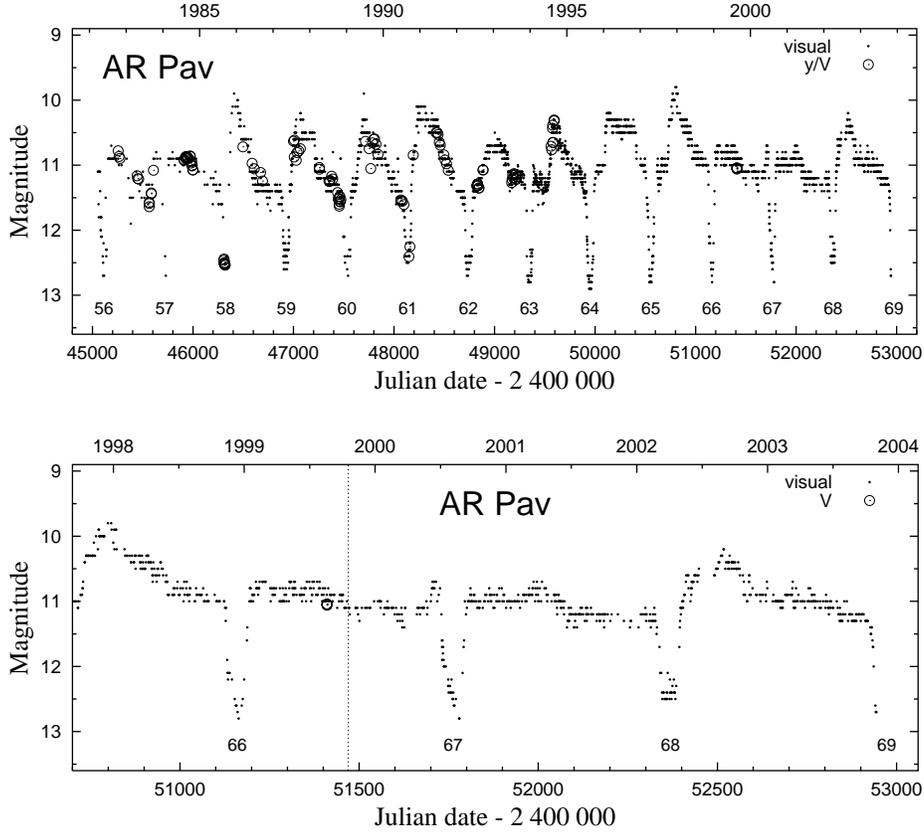

\centering
\begin{center}
\resizebox{\hsize}{!}{\includegraphics[angle=-90]{ar_lc1.epsi}}

\vspace*{6mm}

\resizebox{\hsize}{!}{\includegraphics[angle=-90]{ar_lc2.epsi}}
\caption{
Top: Our visual estimates from 1982.2 to date (made by Albert Jones). 
Bottom: Recent evolution covering a low stage between epochs 
66 and 68.
}
\label{f8}
\end{center}
\end{figure}
%
%

\subsection{AG\,Peg}

We began monitoring this star (HD\,207757) from November 2001. 
The results are summarized in Table~10. 
Stars
HD\,207933 
(SAO\,107460, $V$ = 8.10, $B-V$ = 1.05, $U-B$ = 0.97)
and
HD\,207860 
(SAO\,107453, $V$ = 8.73, $B-V$ = 0.42, spectrum F8),
were used as a comparison and a check star, respectively.

Figure~9 shows our $U,~B,~V$ measurements of AG\,Peg. They show 
a rather complex profile of the LC with respect to, for example, 
recently published photometry of Tomov, Tomova (1998). 
We can see a relatively slow increase in the brightness from 
the beginning of our observations ($\varphi$ = 0.14) to 
a maximum at $\varphi$ = 0.54, then a sudden transition from 
the maximum to a flat minimum in $U$, a plateau at/around 
the maximum in $B$ and a zig-zag variation in $V$.
The last feature of the $V$-LC is similar to that observed 
for CI\,Cyg, which suggests its origin in the cool giant's 
semiregular variability. 
However, our data do not cover the whole orbital cycle and 
only further observations can tell us more about evolution 
of individual components of radiation in the system. 
%
%
\begin{figure}
\centering
\begin{center}
\resizebox{\hsize}{!}{\includegraphics[angle=-90]{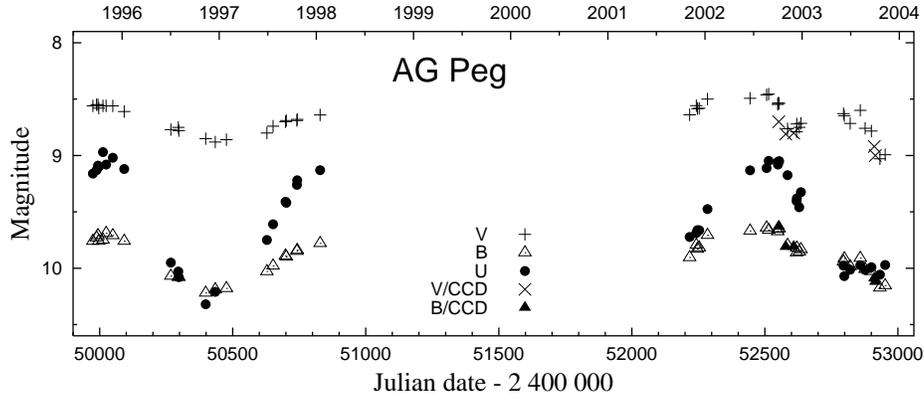}}
\caption{$UBV$ LCs of AG\,Peg. 
Data to 1998 are from Tomov, Tomova (1998). 
}
\label{f9}
\end{center}
\end{figure}
%
%
%
\begin{table}
\scriptsize
\begin{center}
\caption{$U,~B,~V,~R$ observations of AG\,Peg}
\begin{tabular}{lccccccc}
\hline
\hline
Date & JD~24... & Phase$^{\star}$ & $U$ & $B$ & $V$ & $\Delta R$ & Obs \\
\hline
 Nov 03, 01 & 52217.206 & 0.137 &  9.723 &  9.905 &  8.639 & --     & SL \\
 Nov 28, 01 & 52242.203 & 0.167 &  9.688 &  9.795 &  8.558 & -0.040 &SP  \\
 Dec 02, 01 & 52246.242 & 0.172 &  9.663 &  9.831 &  8.584 & --     & SL \\
 Dec 09, 01 & 52253.247 & 0.181 &  9.664 &  9.818 &  8.581 & -0.022 &SP  \\
 Jan 09, 02 & 52284.191 & 0.219 &  9.475 &  9.708 &  8.500 & -0.077 &SP  \\
 Jun 19, 02 & 52444.513 & 0.414 &  9.131 &  9.669 &  8.493 & -0.167 &SP  \\
 Aug 20, 02 & 52506.583 & 0.490 &  9.112 &  9.641 &  8.462 & -0.177 &SP  \\
 Aug 27, 02 & 52514.441 & 0.499 &  9.048 &  9.663 &  8.455 & --     & SL \\
 Sep 30, 02 & 52548.384 & 0.541 &  9.057 &  9.675 &  8.547 & -0.138 &SP  \\
 Oct 01, 02 & 52549.420 & 0.542 &  9.080 &  9.654 &  8.535 & -0.131 &SP  \\
 Oct 04, 02 & 52552.383 & 0.546 & --     &  9.63  &  8.70  & --     & R  \\
 Oct 05, 02 & 52553.477 & 0.547 &  9.050 &  9.648 &  8.537 & -0.126 &SP  \\
 Oct 30, 02 & 52578.273 & 0.577 & --     &  9.81  &  8.81  & --     & R  \\
 Nov 06, 02 & 52585.297 & 0.586 &  9.175 &  9.796 &  8.763 &  0.064 &SP  \\
 Nov 29, 02 & 52608.196 & 0.614 & --     &  9.81  &  8.80  &  --    & R  \\
 Dec 10, 02 & 52619.251 & 0.627 &  9.405 &  9.862 &  8.790 &  0.107 &SP  \\
 Dec 11, 02 & 52620.274 & 0.628 &  9.380 &  9.820 &  8.720 &  0.080 &SP  \\
 Dec 20, 02 & 52629.259 & 0.639 &  9.459 &  9.855 &  8.750 &  0.071 &SP  \\
 Dec 26, 02 & 52635.220 & 0.646 &  9.325 &  9.834 &  8.714 & --     & SL \\
 Jun 04, 03 & 52795.497 & 0.842 &  9.977 &  9.942 &  8.629 & --     & SL \\
 Jun 07, 03 & 52798.450 & 0.845 & 10.070 &  9.914 &  8.647 & -0.009 &SP  \\
 Jun 29, 03 & 52820.463 & 0.872 & 10.013 &  9.984 &  8.715 &  0.074 &SP  \\
 Aug 06, 03 & 52858.458 & 0.919 &  9.974 &  9.911 &  8.599 & -0.043 &SP  \\
 Aug 25, 03 & 52876.560 & 0.941 & 10.016 & 10.010 &  8.758 &  0.074 &SP  \\
 Sep 18, 03 & 52900.514 & 0.970 &  9.991 & 10.028 &  8.783 &  0.126 &SP  \\
 Sep 28, 03 & 52911.371 & 0.983 & --     & 10.09  &  8.92  & --     & R  \\
 Oct 02, 03 & 52915.398 & 0.988 & --     & 10.12  &  9.00  & --     & R  \\
 Oct 19, 03 & 52932.378 & 0.009 & 10.057 & 10.175 &  9.029 &  0.329 &SP  \\
 Nov 08, 03 & 52952.301 & 0.033 &  9.972 & 10.154 &  8.993 & --     & SL \\
\hline
\hline
\end{tabular}
\end{center}
$^{\star}$ $Min = JD\,2\,427\,495.9 + 820.3\times E$
           (Skopal 1998)
\normalsize
\end{table}
%

\subsection{AX\,Per}

The recent measurements of AX\,Per in the $U,B,V,R$ bands are 
given in Table~11 and showed in Fig.~10. 
Star HD\,9839 
(SAO\,22444, $V$ = 7.43, $B-V$ = 1.02, $U-B$ = 0.63) 
and
BD+53\,340 
($V$ = 9.48, $B-V$ = 1.37, $U-B$ = 1.20)
were used as the comparison and check, respectively. 

Figure~10 shows evolution of the $U,~B,~V$ LCs. The wave-like 
profile along the orbital phase indicates that AX\,Per still 
remains at a quiescent phase, which means that the nebular 
emission dominates the near-UV and, in part, optical spectral 
region, mainly around phase 0.5. 
The minimum which occurred at JD~2\,452\,310, was the deepest 
one during the post-outburst period from 1990. On the other hand, 
the minimum indicated by our latest observations was brighter 
by about 0.5\,mag in $U$ and was relatively flat. One can see some 
similarities with the LC evolution of BF\,Cyg. In May 2003 
a 0.5\,mag flare in $B$ an $U$ was detected, followed by a rapid 
decrease of the light. Its phase position ($\varphi \sim$\,0.72) 
and other characteristics are similar to that observed 
in the BF\,Cyg LC (Sect.~3.3). 
%
%
\begin{figure}
\centering
\begin{center}
\resizebox{\hsize}{!}{\includegraphics[angle=-90]{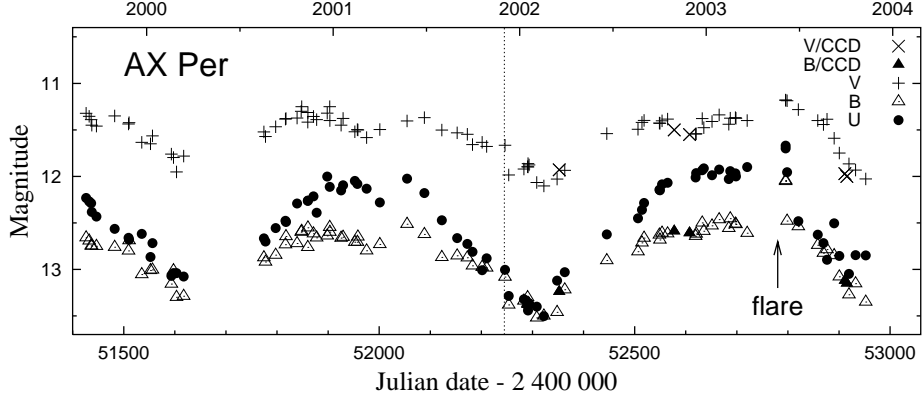}}
\caption{
The $UBV$ LCs of AX\,Per. A small flare developed here 
in May/June 2003. 
}
\label{f10}
\end{center}
\end{figure}
%
\begin{table}
\scriptsize
\begin{center}
\caption{$U,~B,~V,~R$ observations of AX\,Per}
\begin{tabular}{lccccccc}
\hline
\hline
Date & JD~24... & Phase$^{\star}$ & $U$ & $B$ & $V$ & $\Delta R$ & Obs \\
\hline
 Dec 02,  01 & 52246.315 & 0.905 & 13.004 & 13.083 & 11.665 & --     & SL \\
 Dec 09,  01 & 52253.426 & 0.915 & 13.286 & 13.382 & 11.985 &  3.750 &SP  \\
 Jan 08,  02 & 52283.332 & 0.959 & 13.321 & 13.340 & 11.921 &  3.670 &SP  \\
 Jan 15,  02 & 52290.422 & 0.970 & 13.379 & 13.303 & 11.896 &  3.638 &SP  \\
 Jan 16,  02 & 52291.282 & 0.971 & 13.442 & 13.375 & 11.898 &  3.685 &SP  \\
 Jan 16,  02 & 52291.406 & 0.971 & 13.400 & 13.363 & 11.863 & --     & SL \\
 Jan 18,  02 & 52293.287 & 0.974 & 13.364 & 13.373 & 11.877 &  3.581 &SP  \\
 Feb 02,  02 & 52308.290 & 0.996 & 13.40: & 13.52: & 12.065 & --     & SL \\
 Feb 16,  02 & 52322.309 & 0.017 & 13.502 & 13.500 & 12.103 &  3.748 &SP  \\
 Mar 14,  02 & 52348.332 & 0.055 & 13.120 & 13.461 & 12.028 & --     & SL \\
 Mar 18,  02 & 52352.251 & 0.061 & --     & 13.24  & 11.93  & --     & R  \\
 Mar 29,  02 & 52363.295 & 0.077 & 13.030 & 13.219 & 11.935 & --     & SL \\
 Jun 19,  02 & 52445.490 & 0.198 & 12.624 & 12.905 & 11.540 & --     & SL \\
 Aug 20,  02 & 52506.533 & 0.288 & 12.45: & 12.81: & 11.494 &  3.284 &SP  \\
 Aug 27,  02 & 52514.468 & 0.299 & 12.360 & 12.714 & 11.419 & --     & SL \\
 Sep 01,  02 & 52518.595 & 0.305 & 12.286 & 12.663 & 11.398 &  3.150 &SP  \\
 Oct 01,  02 & 52548.501 & 0.349 & 12.144 & 12.629 & 11.427 &  3.166 &SP  \\
 Oct 02,  02 & 52549.549 & 0.351 & 12.149 & 12.684 & 11.429 &  3.171 &SP  \\
 Oct 06,  02 & 52553.570 & 0.357 & 12.086 & 12.610 & 11.402 &  3.164 &SP  \\
 Oct 16,  02 & 52564.421 & 0.373 & 12.069 & 12.617 & 11.384 &  3.146 &SP  \\
 Oct 29,  02 & 52577.487 & 0.392 & --     & 12.59  & 11.50  & --     & R  \\
 Nov 28,  02 & 52607.357 & 0.436 & --     & 12.61  & 11.55  & --     & R  \\
 Dec 10,  02 & 52619.384 & 0.454 & 12.011 & 12.643 & 11.544 &  3.251 &SP  \\
 Dec 11,  02 & 52620.386 & 0.455 & 11.965 & 12.620 & 11.541 &  3.257 &SP  \\
 Dec 23,  02 & 52632.285 & 0.473 & 11.933 & 12.500 & 11.379 &  3.137 &SP  \\
 Dec 26,  02 & 52635.300 & 0.477 & 11.914 & 12.590 & 11.477 & --     & SL \\
 Jan 11,  03 & 52651.369 & 0.501 & 11.987 & 12.531 & 11.410 &  3.116 &SP  \\
 Jan 25,  03 & 52665.430 & 0.521 & 11.926 & 12.465 & 11.338 &  3.046 &SP  \\
 Feb 13,  03 & 52684.307 & 0.549 & 12.029 & 12.559 & 11.440 &  3.114 &SP  \\
 Feb 16,  03 & 52687.288 & 0.553 & 11.943 & 12.455 & 11.374 &  3.060 &SP  \\
 Feb 26,  03 & 52697.306 & 0.568 & 11.967 & 12.512 & 11.368 &  3.067 &SP  \\
 Feb 27,  03 & 52698.302 & 0.570 & 12.001 & 12.517 & 11.376 &  3.066 &SP  \\
 Mar 21,  03 & 52720.314 & 0.602 & 11.900 & 12.611 & 11.400 &  3.068 &SP  \\
 Jun 05,  03 & 52795.515 & 0.713 & 11.699 & 12.054 & 11.179 & --     & SL \\
 Jun 05,  03 & 52795.515 & 0.713 & 11.670 & 12.044 & 11.181 & --     & SL \\
 Jun 07,  03 & 52798.475 & 0.717 & 11.955 & 12.480 & 11.187 &  2.990 &SP  \\
 Jun 30,  03 & 52820.500 & 0.749 & 12.482 & 12.540 & 11.282 &  3.099 &SP  \\
 Aug 07,  03 & 52858.538 & 0.805 & 12.626 & 12.740 & 11.399 &  3.240 &SP  \\
 Aug 17,  03 & 52869.451 & 0.821 & 12.718 & 12.825 & 11.442 &  3.247 &SP  \\
 Aug 25,  03 & 52876.521 & 0.832 & 12.896 & 12.787 & 11.385 &  3.253 &SP  \\
 Sep 08,  03 & 52890.612 & 0.852 & 12.505 & 12.851 & 11.590 &  3.387 &SP  \\
 Sep 18,  03 & 52900.607 & 0.867 & 12.855 & 13.079 & 11.749 &  3.523 &SP  \\
 Sep 28,  03 & 52911.409 & 0.883 & --     & 13.12  & 11.97  & --     & R  \\
 Oct 02,  03 & 52915.463 & 0.889 & --     & 13.15  & 12.00  & --     & R  \\
 Oct 07,  03 & 52919.511 & 0.895 & 13.048 & 13.273 & 11.866 &  3.726 &SP  \\
 Oct 20,  03 & 52932.507 & 0.914 & 12.846 & 13.154 & 11.932 &  3.607 &SP  \\
 Nov 08,  03 & 52952.328 & 0.943 & 12.849 & 13.350 & 12.027 & --     & SL \\
\hline 
\hline
\end{tabular}
\end{center}
 $^{\star}$ $Min = JD\,2\,436\,673.3 + 679.9\times E$
               (Skopal 1991)
\normalsize
\end{table}
%

\subsection{QW\,Sge}

Figure~11 shows our CCD $B,~V,~R_{\rm J},~I_{\rm J}$ photometry. 
We converted our measurements in the $I_{\rm C}$ and $R_{\rm C}$
bands of the Cousins system into the Johnson system according to 
Bessell (1983) by using his transformation equations for 
M giants: 
\begin{equation}
 (V - R)_{\rm J} = 2\times (V - R)_{\rm C} - 0.48, ~~~~~
 (R - I)_{\rm J} = (R - I)_{\rm C} + 0.10 .
\end{equation}
Observations cover the period from 1994.5 to 2003.5 and are given 
in Table~12. Particularly interesting part of the LCs includes an 
active phase, which began at 1997 May and showed two maxima 
in 1997 November and at the beginning of 2000 in all wavelengths. 
The latter is poorly covered, because of a season gap. 
Amplitudes of the first brightening were 
$\Delta B$  = 1.3\,mag, 
$\Delta V$  = 1.4\,mag,
$\Delta R$  = 1.2\,mag, 
and 
$\Delta I$  = 0.8\,mag. 
The brightening in $R$ and $I$ was indicated by our first 
observations at $\sim$\,JD~2\,450\,580, while in $B$ and $V$ 
it started later at about JD~2\,450\,660. These characteristics 
make the brightening of QW\,Sge very different from those 
usually observed for other symbiotic stars - they are significantly 
pronounced at shorter wavelengths (mainly in $U$, see Figs.~1, 2, 7) 
and also are first detected here. 
To reveal the origin of the observed brightenings in the QW\,Sge 
LCs requires a more detailed study. 
%
%
\begin{figure}
\centering
\begin{center}
\resizebox{\hsize}{!}{\includegraphics[angle=-90]{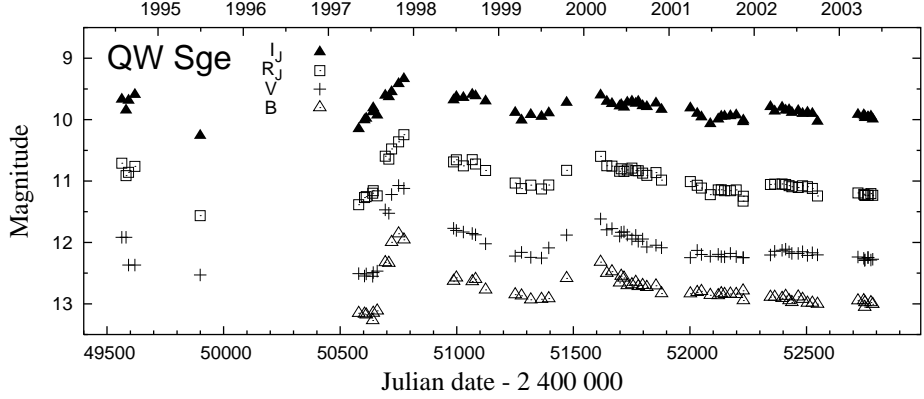}}
\caption{
The CCD $BVRI$ LCs of QW\,Sge.
}
\label{f11}
\end{center}
\end{figure}
%
\begin{table}
\scriptsize
\begin{center}
\caption{$B,~V,~R,~I$ observations of QW\,Sge}
\begin{tabular}{lcccccc}
\hline
\hline
Date        &  JD~24...  &  $I_C$ &  $R_C$ &  $V$   &  $B$   &Obs \\
\hline
 Jul 30, 94 & 49563.561  & 10.136 & 11.073 & --     & --     &PB   \\
 Aug 16, 94 & 49581.448  & 10.211 & 11.174 & 11.917 & --     &PB   \\
 Aug 27, 94 & 49592.406  & 10.304 & 11.373 & 12.368 & --     &PB   \\
 Sep 23, 94 & 49619.423  & 10.258 & 11.325 & --     & --     &PB   \\
 Jun 30, 95 & 49899.494  & 10.605 & 11.806 & 12.529 & --     &PB   \\
 May 10, 97 & 50579.470  & 10.574 & 11.707 & 12.509 & 13.154 &PB   \\
 Jun 04, 97 & 50604.415  & 10.507 & 11.667 & 12.546 & 13.167 &PB   \\
 Jun 11, 97 & 50611.440  & 10.472 & 11.643 & 12.513 & 13.179 &PB   \\
 Jul 09, 97 & 50639.443  & 10.414 & 11.635 & 12.551 & 13.273 &PB   \\
 Jul 11, 97 & 50641.443  & 10.338 & 11.584 & 12.490 & 13.152 &PB   \\
 Jul 28, 97 & 50658.416  & 10.404 & 11.615 & 12.468 & 13.114 &PB   \\
 Sep 01, 97 & 50693.355  &  9.905 & 10.792 & 11.468 & 12.328 &PB   \\
 Sep 16, 97 & 50708.329  &  9.933 & 10.843 & 11.524 & 12.338 &PB   \\
 Sep 28, 97 & 50720.343  &  9.780 & 10.609 & 11.220 & 11.998 &PB   \\
 Oct 28, 97 & 50750.281  &  9.635 & 10.477 & 11.074 & 11.858 &PB   \\
 Nov 20, 97 & 50773.242  &  9.635 & 10.442 & 11.120 & 11.958 &PB   \\
 Jun 21, 98 & 50986.428  & 10.086 & 10.989 & 11.771 & 12.630 &PB   \\
 Jul 03, 98 & 50998.476  & 10.066 & 10.989 & 11.805 & 12.578 &PB   \\
 Aug 02, 98 & 51028.484  & 10.044 & 11.051 & --     & --     &PB   \\
 Sep 09, 98 & 51066.338  & 10.056 & 11.013 & 11.855 & 12.638 &PB   \\
 Sep 23, 98 & 51080.382  & 10.053 & 11.059 & 11.875 & 12.600 &PB   \\
 Nov 05, 98 & 51123.289  & 10.158 & 11.184 & 12.021 & 12.773 &PB   \\
 Mar 13, 99 & 51250.633  & 10.341 & 11.387 & 12.221 & 12.850 &PB   \\
 Apr 09, 99 & 51277.561  & 10.393 & 11.402 & 12.162 & 12.869 &PB   \\
 May 18, 99 & 51317.473  & 10.370 & 11.415 & 12.243 & 12.934 &PB   \\
 Jul 03, 99 & 51363.431  & 10.374 & 11.452 & 12.254 & 12.925 &PB   \\
 Aug 03, 99 & 51394.429  & 10.267 & 11.337 & 12.089 & 12.915 &PB   \\
 Oct 19, 99 & 51471.363  & 10.112 & 11.113 & 11.881 & 12.586 &PB   \\
 Mar 13, 00 & 51616.631  &  9.974 & 10.868 & 11.618 & 12.322 &PB   \\
 Apr 09, 00 & 51643.562  & 10.089 & 11.032 & 11.793 & 12.501 &PB   \\
 Apr 30, 00 & 51665.487  & 10.112 & 11.025 & 11.774 & 12.476 &PB   \\
 Jun 02, 00 & 51698.455  & 10.171 & 11.129 & 11.898 & 12.658 &PB   \\
 Jun 09, 00 & 51705.449  & 10.139 & 11.077 & 11.835 & 12.550 &PB   \\
 Jun 21, 00 & 51717.452  & 10.155 & 11.095 & 11.828 & 12.579 &PB   \\
 Jul 05, 00 & 51731.457  & 10.113 & 11.110 & 11.881 & 12.702 &PB   \\
 Jul 25, 00 & 51751.472  & 10.142 & 11.127 & 11.944 & 12.686 &PB   \\
 Aug 10, 00 & 51767.413  & 10.126 & 11.117 & 11.884 & 12.649 &PB   \\
 Aug 21, 00 & 51778.378  & 10.142 & 11.166 & 11.980 & 12.717 &PB   \\
 Sep 09, 00 & 51797.336  & 10.175 & 11.167 & 11.944 & 12.700 &PB   \\
 Sep 27, 00 & 51815.308  & 10.236 & 11.251 & 12.074 & 12.731 &PB   \\
 Nov 05, 00 & 51854.274  & 10.187 & 11.214 & 12.045 & 12.706 &PB   \\
 Nov 29, 00 & 51878.230  & 10.249 & 11.296 & 12.086 & 12.832 &PB   \\
 Apr 02, 01 & 52001.563  & 10.293 & 11.388 & 12.247 & 12.836 &PB   \\
 May 01, 01 & 52031.478  & 10.294 & 11.359 & 12.130 & 12.814 &PB   \\
 May 19, 01 & 52049.440  & 10.362 & 11.416 & 12.197 & 12.797 &PB   \\
 Jun 26, 01 & 52087.427  & 10.433 & 11.484 & 12.227 & 12.861 &PB   \\
 Jul 30, 01 & 52121.447  & 10.381 & 11.430 & 12.195 & 12.865 &PB   \\
 Aug 12, 01 & 52134.418  & 10.355 & 11.446 & 12.230 & 12.834 &PB   \\
 Aug 26, 01 & 52148.361  & 10.346 & 11.455 & 12.234 & 12.847 &PB   \\
 Sep 20, 01 & 52173.321  & 10.313 & 11.428 & 12.179 & 12.839 &PB   \\
 Oct 15, 01 & 52198.296  & 10.331 & 11.447 & 12.230 & 12.845 &PB   \\
 Nov 14, 01 & 52228.264  & 10.359 & 11.547 & --     & --     &PB   \\
 Nov 15, 01 & 52229.247  & 10.372 & 11.507 & 12.247 & 12.787 &PB   \\
 Mar 12, 02 & 52345.618  & 10.227 & 11.391 & 12.205 & 12.947 &PB   \\
 Mar 30, 02 & 52363.574  & 10.274 & 11.357 & 12.145 & 12.891 &PB   \\
 May 01, 02 & 52395.511  & 10.203 & 11.347 & 12.127 & 12.892 &PB   \\
 May 15, 02 & 52410.454  & 10.240 & 11.343 & 12.111 & 12.910 &PB   \\
 May 30, 02 & 52425.433  & 10.243 & 11.375 & 12.154 & 12.876 &PB   \\
 Jun 11, 02 & 52437.455  & 10.292 & 11.396 & 12.180 & 12.947 &PB   \\
 Jul 08, 02 & 52464.433  & 10.254 & 11.403 & 12.180 & 12.977 &PB   \\
 Jul 27, 02 & 52483.407  & 10.302 & 11.375 & 12.152 & 12.886 &PB   \\
\hline                                                        
\end{tabular}                                                 
\end{center}                                                  
\end{table}                                           
\addtocounter{table}{-1}                              
\begin{table}                                         
\scriptsize                                           
\begin{center}                                        
\caption{Continued}
\begin{tabular}{cccccccc}
\hline
Date        &  JD~24...  &  $I_C$ &  $R_C$ &  $V$   &  $B$   &Obs \\
\hline
 Aug 17, 02 & 52504.369  & 10.305 & 11.412 & 12.211 & 12.925 &PB   \\
 Sep 07, 02 & 52525.342  & 10.288 & 11.405 & 12.169 & 12.987 &PB   \\
 Sep 29, 02 & 52547.290  & 10.372 & 11.483 & 12.203 & 12.987 &PB   \\
 Mar 22, 03 & 52720.597  & 10.303 & 11.476 & 12.238 & 13.006 &PB   \\
 Apr 16, 03 & 52745.533  & 10.309 & 11.497 & 12.259 & 12.949 &PB   \\
 Apr 20, 03 & 52749.533  & 10.339 & 11.519 & 12.292 & 12.948 &PB   \\
 Apr 21, 03 & 52750.533  & 10.356 & 11.519 & 12.284 & 13.054 &PB   \\
 May 03, 03 & 52763.492  & 10.329 & 11.499 & 12.246 & 12.979 &PB   \\
 May 16, 03 & 52776.455  & 10.350 & 11.505 & 12.286 & 12.980 &PB   \\
 May 24, 03 & 52784.456  & 10.377 & 11.513 & 12.274 & 13.007 &PB   \\
%
\hline                                                        
\hline                                                        
\end{tabular}                                                 
\end{center}                                                  
\normalsize                                           
\end{table}                                           
%

\subsection{IV\,Vir}

The $U,B,V$ measurements of IV\,Vir (BD-21\,3873) are listed in 
Table~13. Observations were carried out at the San Pedro 
Observatory in 2003 April at the orbital phase 0.25. 
The star 
HD\,124991 (BD-21\,3877;  $V$ = 8.072, $B-V$ = 1.048, $U-B$ = 0.730)
was used as the comparison. 

The brightness of IV\,Vir is very close to that obtained by us at 
the orbital phase 0.84 (Table~9 of S+02). This confirms the 
wave variability of this quiet symbiotic, which produces a roughly 
symmetrical LC. Here, the measurements at phases 0.84 and 0.25 lie 
on the descending and the ascending branch of the broad minimum 
(see Fig.~9 of S+02).
%
%
\begin{table}[!ht]
\scriptsize
\begin{center}
\caption{$U,B,V$ observations of IV\,Vir}
\begin{tabular}{lcccccc}
\hline
\hline
Date & JD~24... & Phase$^{\star}$ & $U$ &   $B$   &   $V$  & Obs \\
\hline
 Apr 20, 03 & 52749.843 & 0.256 & 12.956 & 12.139 & 10.689 & M  \\
 Apr 20, 03 & 52749.844 & 0.256 & 12.979 & 12.078 & 10.698 & M  \\
 Apr 21, 03 & 52750.852 & 0.260 & 12.924 & 12.139 & 10.668 & M  \\
 Apr 21, 03 & 52750.855 & 0.260 & --     & 12.131 & 10.696 & M  \\
\hline
\hline
\end{tabular}
\end{center}
$^{\star}$ $JD_{\rm sp. conj.} = 2\,449\,016.9 + 281.6\times E$
           (Smith {\it et al.} 1997)
\normalsize
\end{table}
%

\subsection{V934\,Her}

This peculiar M giant (HD\,154791) is the only optical counterpart 
of a hard X-ray source (4U1700+24) and can be classified as a LMXB 
(Gaudenzi, Polcaro 1999). This object was included in our 
observing programme to indicate possible optical variability 
as a response to an X-ray outburst, which peaked between
August and September 2002 (Galloway, private communication).  

Our photometric observations are summarized in Table~14. 
Star HD\,155104
(SAO\,84873, $V$ = 6.85, $B-V$ = 0.13, $U-B$ = 0.10, spectrum B5)
and
GCS:02060-00124
($V$ = 10.06, $B-V$ = 0.73, spectrum F1\,V)
were used as the comparison and check, respectively.

Our observations did not show any larger variation in the optical. 
We detected only a small brightening by about 0.12\,mag in $U$ and 
by 0.06 -- 0.07\,mag in other bands between observations made on 
13-16 and 25-27 February 2003 (see Table~14). These values are 
well above the uncertainties of the given means of all individual 
measurements made during the night. We note that we measured 
V934\,Her for about 1 hour to obtain 12 -- 25 individual points 
(i.e. the magnitude difference between the target and 
the comparison star). 

The recent studies (e.g. Galloway {\it et al.} 2002; Masetti 
{\it et al.} 2002) suggest that the object consists of a wide
binary system ($P_{\rm orb}$ = 400\,d), in which a neutron star 
accretes matter from the wind of a M-type giant star. Their 
observations showed that the X-ray luminosity ranges from about 
$2\times 10^{32}$ to $1\times 10^{34}$\, erg s$^{-1}$. 
Qualitatively, the very small and/or negligible response to 
the X-ray outburst in the optical wavelengths (also 
Tomasella {\it et al.} 1997) suggests too little amount of 
the circumstellar material in the system, which implies a small 
mass loss rate from the giant, and thus also a small accretion rate. 
If we assume that the only energy source is accretion, a neutron 
star has to accrete matter at the rate of 
$\sim\,1\times 10^{-12}\,M_{\sun}$\,yr$^{-1}$
to 
$\sim\,1\times 10^{-10}\,M_{\sun}$\,yr$^{-1}$ 
to balance the observed X-ray luminosities (we adopted 
$M_{\rm acc} = 1.4\,M_{\sun}$ and $R_{\rm acc}$ = 100\,km). 
In the opposite case, a variability 
in the X-ray radiation would imply a more significant variation 
in the near-UV and optical continuum as well as in highly ionized 
emission lines (e.g. He\I\I\,1640) as it is, generally, observed 
for classical symbiotic stars. 
%
\begin{table}
\scriptsize
\begin{center}
\caption{$U,~B,~V,~R$ observations of V934\,Her}
\begin{tabular}{lcccccc}
\hline
\hline
    Date    &  JD~24... &   $U$  &  $B$   &   $V$  &$\Delta R$ & Obs \\
\hline
 Feb 14, 03 & 52684.609 & 11.030 &  9.255 &  7.623 & -0.052 &SP  \\
 Feb 17, 03 & 52687.569 & 11.110 &  9.252 &  7.671 & -0.049 &SP  \\
 Feb 26, 03 & 52696.553 & 10.983 &  9.197 &  7.593 & -0.112 &SP  \\
 Feb 27, 03 & 52697.557 & 10.989 &  9.188 &  7.582 & -0.123 &SP  \\
 Feb 28, 03 & 52698.547 & 10.983 &  9.184 &  7.580 & -0.121 &SP  \\
\hline
\hline
\end{tabular}
\end{center}
\normalsize
\end{table}
%

\acknowledgements
This research was supported by a grant of the Slovak Academy of 
Sciences No. 2/4014/4, by allocation of San Pedro M\'artir 
observing time, from the Czech-Mexican project ME 402/2000 and 
by the research plan J13/98: 113200004 Investigation of the 
Earth and the Universe. Authors thank the referee, Dr. Sergei 
Shugarov, for helpful comments.

\end{document}